\documentclass[floats,floatfix,twocolumn,superscriptaddress,nofootinbib,showpacs]{revtex4-1}

\pdfoutput=1
\usepackage{amsfonts,amssymb,amsmath}
\usepackage[usenames,dvipsnames]{xcolor}
\definecolor{292}{rgb}{0.3843, 0.6588, 0.8980}
\usepackage[unicode, colorlinks=true, 
	linkcolor=blue, 
	urlcolor=blue, 
	citecolor=292]{hyperref}
\usepackage[all]{hypcap}
\usepackage{graphicx}
\usepackage{xspace}
\usepackage{tikz}
\usepackage{float}
\usetikzlibrary{shapes,arrows}
\usetikzlibrary{matrix}
\usetikzlibrary{shapes.multipart}
\usetikzlibrary{er,positioning}
\usepackage[normalem]{ulem} 
\usepackage{url}
\usepackage{color,soul}
\usepackage[caption=false]{subfig}
\usepackage[T1]{fontenc}
\usepackage[utf8]{inputenc}
\usepackage{microtype}
\newcommand{\roughly}{\mathchar"5218\relax} 

\usepackage{orcidlink}

\newcommand{\Caltech}{\affiliation{Theoretical Astrophysics,
    Walter Burke Institute for Theoretical Physics,\\
    California Institute of Technology, Pasadena, California 91125, USA}}
\newcommand{\Cornell}{\affiliation{Cornell Center for Astrophysics and
    Planetary Science, Cornell University, Ithaca, New York 14853, USA}}

\makeatletter
\newsavebox\myboxA
\newsavebox\myboxB
\newlength\mylenA

\newcommand*\xoverline[2][0.75]{%
	\sbox{\myboxA}{$\m@th#2$}%
	\setbox\myboxB\null
	\ht\myboxB=\ht\myboxA%
	\dp\myboxB=\dp\myboxA%
	\wd\myboxB=#1\wd\myboxA
	\sbox\myboxB{$\m@th\overline{\copy\myboxB}$}
	\setlength\mylenA{\the\wd\myboxA}
	\addtolength\mylenA{-\the\wd\myboxB}%
	\ifdim\wd\myboxB<\wd\myboxA%
	\rlap{\hskip 0.5\mylenA\usebox\myboxB}{\usebox\myboxA}%
	\else
	\hskip -0.5\mylenA\rlap{\usebox\myboxA}{\hskip 0.5\mylenA\usebox\myboxB}%
	\fi}
\makeatother

\begin{document}

\title{Computation of Displacement and Spin Gravitational Memory in Numerical Relativity}

\author{Keefe Mitman \orcidlink{0000-0003-0276-3856}}
\email{kmitman@caltech.edu}
\Caltech
\author{Jordan Moxon}
\Caltech
\author{Mark A. Scheel}
\Caltech
\author{Saul A. Teukolsky}
\Caltech\Cornell
\author{Michael Boyle}
\Cornell
\author{Nils Deppe}
\Cornell
\author{Lawrence E. Kidder}
\Cornell
\author{William Throwe}
\Cornell

\date{\today}

\begin{abstract}
\noindent We present the first numerical relativity waveforms for binary black hole mergers produced using spectral methods that show both the displacement and the spin memory effects. Explicitly, we use the SXS Collaboration's $\texttt{SpEC}$ code to run a Cauchy evolution of a binary black hole merger and then extract the gravitational wave strain using $\texttt{SpECTRE}$'s version of a Cauchy-characteristic extraction. We find that we can accurately resolve the strain's traditional $m=0$ memory modes and some of the $m\not=0$ oscillatory memory modes that have previously only been theorized. We also perform a separate calculation of the memory using equations for the Bondi-Metzner-Sachs charges as well as the energy and angular momentum fluxes at asymptotic infinity. Our new calculation uses only the gravitational wave strain and two of the Weyl scalars at infinity. Also, this computation shows that the memory modes can be understood as a combination of a memory signal throughout the binary's inspiral and merger phases, and a quasinormal mode signal near the ringdown phase. Additionally, we find that the magnetic memory, up to numerical error, is indeed zero as previously conjectured. Lastly, we find that signal-to-noise ratios of memory for LIGO, the Einstein Telescope (ET), and the Laser Interferometer Space Antenna (LISA) with these new waveforms and new memory calculation are larger than previous expectations based on post-Newtonian or Minimal Waveform models.

\end{abstract}

\maketitle

\section{Introduction}
\label{sec:introduction}

As has been understood since the early 1970s~\cite{Zeldovich_1974, Thorne_1987, Christodoulou_1991, Thorne_1992}, when gravitational waves (GWs) pass through the arms of a GW detector, a persistent physical change to the corresponding region of spacetime is induced as a result of the transient radiation. Originally, this effect, which is referred to as the \emph{memory effect} or just \emph{memory}, was found by studying the fly-by behavior of two compact astrophysical objects that travel to asymptotic infinity as $t\rightarrow+\infty$ on timelike paths~\cite{Zeldovich_1974}. Later, it was realized that the memory effect also occurs when null radiation travels to asymptotic null infinity as $r,t\rightarrow+\infty$ at a fixed Bondi time $u\equiv t-r$~\cite{Christodoulou_1991}. Originally, these two unique contributions to memory were called \emph{linear memory} and \emph{nonlinear memory}\footnote{Also known as \emph{Christodoulou memory}~\cite{Christodoulou_1991, Thorne_1992}.} because of the order of the metric's perturbative expansion that was used to calculate each of the independent memory contributions.

Recently, the memory effect was realized to be the element needed to extend the Poincar\'e conservation laws to the infinite number of proper Bondi-Metzner-Sachs (BMS) conservation laws~\cite{Strominger_2014, Pasterski_2016, Nichols_2018, compre2019poincar}, which correspond to the various BMS and extended BMS transformations~\cite{doi:10.1098/rspa.1962.0161, doi:10.1098/rspa.1962.0206, Barnich_2010, de_Boer_2003, banks2003critique, barnich2011supertranslations, Kapec_2014, Kapec_2017, He_2017}, i.e., supertranslations, superrotations, and superboosts.\footnote{Formally, superrotations and superboosts, which are the two types of super-Lorentz transformations, can be realized as the \(|m|\geq2\) elements of the Virasoro algebra $\big($an extension of the more common M\"obius transformations, i.e., $PL(2,\mathbb{C})\big)$, just as supertranslations can be viewed as the \(l\geq2\) spherical harmonics. These super-Lorentz transformations, though, which form the extended BMS group, do not preserve asymptotic flatness.} Unlike the ten Poincar\'e conservation laws, which equate the change in the Poincar\'e charges to the corresponding energy and momentum fluxes, the BMS conservation laws state that the change in the BMS charges minus the corresponding fluxes\footnote{Often, the BMS conservation law is written as
\begin{align}
\text{Change in BMS charges} - \text{BMS fluxes} = 0,
\end{align}
where the ``BMS flux'' is understood to have two contributions: ``hard'' and ``soft,'' with the hard contribution being the flux in Eq.~\eqref{eq:BMSLaw} and the soft contribution being the memory in Eq.~\eqref{eq:BMSLaw}.} is exactly the memory effect, i.e.,
\begin{align}
\label{eq:BMSLaw}
\text{Change in BMS charges}-\text{BMS fluxes}=\text{Memory}.
\end{align}

Early studies of gravitational memory focused on the type of memory corresponding to supertranslations and supermomentum, which is called \emph{displacement memory}. We follow \cite{Pasterski_2016} and~\cite{Nichols_2018} and refer to the other memory effects, which are related to superrotations and superboosts, as the \emph{spin} and the \emph{center-of-mass} (CM) memory effects. While the displacement memory is the most prominent in the strain of a gravitational wave, the spin and CM memory effects can most easily be noticed in the time integral of the strain. Physically, displacement memory is related to a change in a GW detector's arm length~\cite{Zeldovich_1974, Thorne_1987, Christodoulou_1991, Thorne_1992}, while the spin memory relates to the relative time delay that would be acquired by counter-orbiting objects, e.g., the particle beams in the Large Hadron Collider or a freely falling Sagnac interferometer~\cite{Pasterski_2016}. The CM memory, by contrast, corresponds to the relative time delay that would be acquired by objects on antiparallel paths~\cite{Nichols_2018}. As an example, for two particles bouncing back and forth in a Fabry-Perot cavity, if a gravitational wave propagates at an angle through the cavity, then the particles will acquire a relative time delay given by the CM memory.

Furthermore, because the various memory effects are now known to be calculable from BMS flux-balance laws, both of the previous classifications of linear and nonlinear contributions have been renamed to be more indicative of what they represent. Instead, the two contributions to each of the three memory effects are now referred to as the \emph{ordinary memory} and the \emph{null memory}. Moreover, the modern nomenclature also avoids potential confusion about which types of terms should be included in each memory effect because whether a particular effect appears linearly or nonlinearly varies with the perturbation theory that is being considered~\cite{Bieri_2014}. As one might expect, for the most common sources of observable GW radiation, i.e., binary black holes (BBHs), the displacement memory is the most prominent, followed by the spin memory, and then the center-of-mass memory~\cite{Nichols_2018}.

Over the past few years, there have been many studies of whether current or future GW detectors could measure the displacement and the spin memory effects~\cite{Favata_2009_PN, Nichols_2017, Lasky_2016, Talbot_2018, Boersma_2020, Hubner_2020}. These previous studies, however, used approximations of the  memory since earlier calculations of the memory in a BBH merger have, until now, been incomplete. For one, the waveforms produced by numerical simulations using extrapolation techniques have been unable to resolve the primary $m=0$ memory modes and have also failed to produce the expected memory in certain oscillatory $m\not=0$ memory modes.\footnote{While the strain $(2,0)$ mode, which is the primary contributor to the displacement memory, has been resolved previously~\cite{Pollney_2011}, the code used in this work was much more computationally expensive and thus could not easily run longer simulations required to accurately resolve the other memory effects.} Apart from this, previous calculations of memory have used post-Newtonian (PN) approximations or have tried to compute an effective memory using the available numerical waveforms through various kinds of postprocessing techniques~\cite{Favata_2009,Favata_2009_PN,Favata_2010,Talbot_2018}.

So far, PN approximations have been computed for the modes contributing to the displacement memory through 3PN order, through 2.5PN order for the spin memory, and even through 3PN for the CM memory~\cite{Favata_2009_PN,Nichols_2017,Nichols_2018}. However, the memory effect is predominantly accumulated during the merger phase of a BBH coalescence, in which most of the system's energy and angular momentum are radiated by GWs. Because PN theory cannot capture the merger phase of a BBH coalescence, we must instead use numerical relativity (NR) simulations to calculate the displacement, spin, and CM memory effects.

As already mentioned, previous numerical relativity simulations have been unable to extract the three unique memory effects for a variety of reasons~\cite{Favata_2009_PN}. For one, numerical relativity simulations of BBH mergers typically compute the strain on concentric finite-radius spheres and then extrapolate the strain to future null infinity using a collection of fits. While this procedure is adequate for computing the main strain modes, it unfortunately does not produce waveforms that accurately resolve the modes responsible for illustrating the various memory effects. As a result, even though approximate calculations of the memory in the strain can be performed using waveforms that have been computed thus far, they will nonetheless be incomplete since they fail to include the next-order memory contributions from the fluxes induced by the memory modes themselves. Furthermore, many of these postprocessing computations of the memory use only the primary waveform modes---often just the $(2,2)$ mode---instead of every mode. This is because, before this work, there has not been a method for fully computing the memory from every mode of a waveform.\footnote{In~\cite{Talbot_2018} a procedure using the result of~\cite{Thorne_1992} was presented for computing just the displacement memory using all of the modes of a strain waveform. However, this method was only used on extrapolated waveforms, which exhibit no displacement memory, and thus fails to accurately capture the ``memory of the memory'', i.e., the memory induced by the memory modes.} 

As a part of this study, we present the first successful resolution of the modes that contain memory by using the Simulating eXtreme Spacetimes (SXS) Collaboration's older and newer codes, $\texttt{SpEC}$~\cite{Boyle_2019} and $\texttt{SpECTRE}$~\cite{SpECTRE}. Explicitly, we use Cauchy-characteristic extraction (CCE) to evolve a world tube produced by a Cauchy evolution to asymptotic infinity, where we extract many observables, most importantly the strain~\cite{moxon2020improved}. With CCE, we find that we can resolve many of the $m=0$ and $m\not=0$ modes that contribute to the displacement and spin memories. Through this, we observe that not only do CCE waveforms surpass extrapolated waveforms in terms of resolving the displacement memory, but they also exhibit a spin memory that is roughly twice as much as what is seen in the extrapolated waveforms~\cite{Nichols_2017}. Furthermore, we compare the displacement and spin memory modes to the memory computed from the numerical waveforms using the new memory equations presented in this paper. We find that the two agree exceptionally well, which implies that the CCE waveforms obey the BMS flux-balance laws to a rather high degree of accuracy. We also briefly discuss the CM memory's formulation in Sec.~\ref{sec:cmmemorycomputation} and its presence in our numerical results in Appendix~\ref{a:cmmem}.

\subsection{Overview}
\label{sec:overview}
We organize our computations and results as follows. Using Einstein's field equations, we compute expressions for the displacement and spin memory in Secs.~\ref{sec:bondiframework} and~\ref{sec:memorycomputation}, which are valid in asymptotically flat spacetimes. Moreover, we write these expressions in terms of the observables that are explicitly produced by SXS's CCE. We also provide a few brief comments on the CM memory in Sec.~\ref{sec:cmmemorycomputation}, but not a complete mathematical expression. Following this, in Sec.~\ref{sec:ccevsextrapolation}, we describe certain aspects of CCE and outline the choices that we make to produce memory results that agree with post-Newtonian theory. Note, we explore the features of CCE further in Sec.~\ref{sec:cce}. Continuing to our numerical results, in Sec.~\ref{sec:fluxbalance}, we then illustrate how well our extracted observables comply with the BMS flux-balance laws that we compute in Sec.~\ref{sec:memorycomputation}. Next, in Secs.~\ref{sec:ememory},~\ref{sec:bmemory}, and~\ref{sec:spinmemory}, we present the results for five numerical simulations covering combinations of equal and unequal masses, spinning and nonspinning, and precessing and nonprecessing, whose parameters are outlined in the introduction of Sec.~\ref{sec:results}. We not only show the success of CCE in resolving the modes that express memory effects, but also compare them to the memory that is expected according to our calculations in Sec.~\ref{sec:memorycomputation}. Furthermore, in Sec.~\ref{sec:qnms}, we show that during ringdown, the most prominent memory modes can be accurately modeled as a sum of the null memory contribution and the corresponding quasinormal modes (QNMs) of the remnant BH. Finally, in Sec.~\ref{sec:snrs}, with these results we then compute signal-to-noise ratios (SNRs) for LIGO, ET, and LISA and thus provide estimates on the measurability of both the displacement and the spin memory effects. We also provide computations of the Bondi mass aspect and the Bondi angular momentum aspect in Appendixes~\ref{a:bondimass} and~\ref{a:bondiangmom} in terms of the strain and the Weyl scalars $\Psi_{2}$ and $\Psi_{1}$. Appendix~\ref{a:cmmem} gives an informal presentation of a mode of the strain that exhibits the CM memory effect.

\subsection{Conventions}
\label{sec:conventions}
We set $c=G=1$. When working with complex dyads, following the work of Moxon \emph{et al.}~\cite{moxon2020improved}, we use
\begin{align}
q_{A}=-(1,i\sin\theta)\text{ and }q^{A}=-(1,i\csc\theta),
\end{align}
and denote the round metric on the two-sphere as $q_{AB}$. The complex dyad obeys the following properties
\begin{align}
q_{A}q^{A}=0,\,q_{A}\bar{q}^{A}=2,\,q_{AB}=\frac{1}{2}(q_{A}\bar{q}_{B}+\bar{q}_{A}q_{B}).
\end{align}
We built spin-weighted fields with the dyads as follows. For a tensor field $W_{A\cdots D}$, the function
\begin{align}
W=W_{A\cdots BC\cdots D}q^{A}\cdots q^{B}\bar{q}^{C}\cdots\bar{q}^{D}
\end{align}
with $m$ factors of $q$ and $n$ factors of $\bar{q}$ has spin-weight $s=m-n$. We raise and lower spins using the differential spin-weight operators $\eth$ and $\bar{\eth}$,
\begin{subequations}
\begin{align}
\eth W&=(D_{E}W_{A\cdots BC\cdots D})q^{A}\cdots q^{B}\bar{q}^{C}\cdots\bar{q}^{D}q^{E},\\
\bar{\eth}W&=(D_{E}W_{A\cdots BC\cdots D})q^{A}\cdots q^{B}\bar{q}^{C}\dots\bar{q}^{D}\bar{q}^{E}.
\end{align}
\end{subequations}
Here, $D_{A}$ is the covariant derivative on the two-sphere. The $\eth$ and $\bar{\eth}$ operators in spherical coordinates are then
\begin{subequations}
\begin{align}
\eth W(\theta,\phi)&=-(\sin\theta)^{+s}(\partial_{\theta}+i\csc\theta\partial_{\phi})\nonumber\\
&\phantom{=.-(\sin(\theta))^{s}}\left[(\sin\theta)^{-s}W(\theta,\phi)\right],\\
\bar{\eth} W(\theta,\phi)&=-(\sin\theta)^{-s}(\partial_{\theta}-i\csc\theta\partial_{\phi})\nonumber\\
&\phantom{=.-(\sin(\theta))^{s}}\left[(\sin\theta)^{+s}W(\theta,\phi)\right].
\end{align}
\end{subequations}
Thus, when acting on spin-weighted spherical harmonics, these operators produce
\begin{subequations}
\begin{align}
\eth(\phantom{}_{s}Y_{\ell m})&=+\sqrt{(\ell-s)(\ell+s+1)}_{s+1}Y_{\ell m},\\
\bar{\eth}(\phantom{}_{s}Y_{\ell m})&=-\sqrt{(\ell+s)(\ell-s+1)}_{s-1}Y_{\ell m}.
\end{align}
\end{subequations}
As a result, for $f(\theta,\phi)$ an arbitrary spin-weight 0 function, the spherical Laplacian $D^{2}$ is then given by
\begin{align}
D^{2}f(\theta,\phi)=\eth\bar{\eth}f(\theta,\phi)=\bar{\eth}\eth f(\theta,\phi).
\end{align}
Last, for our comparisons to PN computations, we use the polarization convention that coincides with Kidder~\cite{Kidder_2008}, rather than Blanchet~\cite{Blanchet_2008}, since most PN calculations of the memory make this choice as well~\cite{Favata_2009_PN, Nichols_2017}.

\section{Description of Memory}
\label{sec:memoryintroduction}

We now review the mathematical formulation of the memory effects and extend previous results to be more relevant for calculations in numerical relativity.

\subsection{Bondi Framework}
\label{sec:bondiframework}

We begin by reviewing a few of Einstein's equations for the asymptotically flat Bondi-Sachs metric to obtain relationships between conserved charge quantities and memory-contributing terms. We closely follow the work of Flanagan and Nichols~\cite{Flanagan_2017}, but we only consider a vacuum spacetime. We extend their results by computing the memory contribution to the gravitational wave strain, i.e., the quantity that is extracted in numerical relativity and currently measured by GW detectors.

Consider retarded Bondi coordinates, $(u,r,\theta^{1},\theta^{2})$, near future null infinity, where $u\equiv t-r$. For such a system, the metric of arbitrary asymptotically flat spacetimes can be written in the form
\begin{align}
ds^2&=-Ue^{2\beta}du^2-2e^{2\beta}dudr\nonumber\\
&\phantom{=.}+r^2\gamma_{AB}(d\theta^{A}-\mathcal{U}^{A}du)(d\theta^{B}-\mathcal{U}^{B}du),
\end{align}
where $A,B\in\{1,2\}$ are coordinates on the two-sphere, and $U$, $\beta$, $\mathcal{U}^{A}$, and $\gamma_{AB}$ are functions of $u$, $r$, and $\theta^{A}$. Here we apply the four gauge conditions
\begin{align}\label{eq:bondi_gauge_conditions}
g_{rr}=0,\,g_{rA}=0,\text{ and }\det(\gamma_{AB})=\det(q_{AB}),
\end{align}
where \(g_{\mu\nu}\) is the metric of four-dimensional spacetime.
We now expand these metric functions as series in $1/r$ to relevant orders, which gives
\begin{align}
U&=1-\frac{2m}{r}-\frac{2\mathcal{M}}{r^2}+\mathcal{O}(r^{-3}),\\
\beta&=\frac{\beta_{0}}{r}+\frac{\beta_{1}}{r^2}+\frac{\beta_{2}}{r^3}+\mathcal{O}(r^{-4}),\\
\label{eq:metricUterm}
\mathcal{U}^{A}&=\frac{U^{A}}{r^2}+\frac{1}{r^3}\Big[-\frac{2}{3}N^{A}+\frac{1}{16}D^{A}(C_{BC}C^{BC})\nonumber\\
&\phantom{=.\frac{U^{A}}{r^2}+\frac{1}{r^3}\Big[}+\frac{1}{2}C^{AB}D^{C}C_{BC}\Big]+\mathcal{O}(r^{-4}),\\
\gamma_{AB}&=
q_{AB}+\frac{C_{AB}}{r}+\frac{D_{AB}}{r^2}+\frac{E_{AB}}{r^3}+\mathcal{O}(r^{-4}),
\end{align}
where the various coefficients on the right-hand sides are functions of $(u,\theta^{A})$ only, and $q_{AB}(\theta^{A})$ is the metric on the two-sphere, i.e., $q_{AB}(\theta,\phi)=d\theta^2+\sin^2\theta\,d\phi^2$ in ordinary spherical coordinates. The three most important functions above are: the Bondi mass aspect $m$, the Bondi angular momentum aspect $N^A$, and the shear tensor $C_{AB}$, whose retarded time derivative is the Bondi news tensor $N_{AB}\equiv\partial_{u}C_{AB}$. The Bondi mass aspect is related to the supermomentum charge while the angular momentum, once a few extra terms are included,\footnote{Extra terms are needed because the angular momentum aspect cannot explicitly be related to one of the conserved BMS charges; see Sec.~\ref{sec:bondiframework} for a further explanation.} corresponds to the super-Lorentz charges~\cite{compre2019poincar}. Applying the gauge conditions in Eq.~\eqref{eq:bondi_gauge_conditions} produces the constraints
\begin{align}
q^{AB}C_{AB}&=0,\\
D_{AB}&=\frac{1}{4}q_{AB}C_{CD}C^{CD}+\mathcal{D}_{AB},\\
E_{AB}&=\frac{1}{2}q_{AB}C_{CD}\mathcal{D}^{CD}+\mathcal{E}_{AB},
\end{align}
where $\mathcal{D}_{AB}$ and $\mathcal{E}_{AB}$ are two arbitrary traceless tensors.

Finally, we consider Einstein's equations. By computing the $\mathcal{O}(1/r^2)$ terms of the $uu$ part of the evolution equation for the Bondi mass aspect, we find
\begin{align}
\label{eq:bondimassevo}
\dot{m}&=-\frac{1}{8}N_{AB}N^{AB}+\frac{1}{4}D^{A}D^{B}N_{AB}.
\end{align}
Equation~\eqref{eq:bondimassevo} is identical to the central result of~\cite{doi:10.1098/rspa.1962.0161}, which outlines the link between a system's news and mass loss.\footnote{The reason why the $D^{A}D^{B}N_{AB}$ term was not important in~\cite{doi:10.1098/rspa.1962.0161} is because they integrated their version of Eq.~\eqref{eq:bondimassevo} over the sphere, which kills this term because its $\ell=0,1$ modes are zero.} If we integrate and reorder this equation, we obtain
\begin{align}
\label{eq:bondimassevo_null}
\frac{1}{4}D^{A}D^{B}C_{AB}=m+4\pi\mathcal{E},
\end{align}
where
\begin{align}
\mathcal{E}=\frac{1}{32\pi}\int N_{AB}N^{AB}du
\end{align}
is just the energy that is radiated per unit solid angle. Equation~\eqref{eq:bondimassevo_null} represents one of the two BMS flux-balance laws that we will examine. The first term corresponds to the memory appearing in the shear. The second term, which relates to the ordinary memory contribution, can be understood as the change in a BMS charge---specifically, the supermomentum charge. The third term, which can be viewed as the null memory contribution, is a flux---specifically, an energy flux. We now repeat the calculation performed above, but for the angular momentum aspect.

Computing the $\mathcal{O}(1/r^2)$ terms of the $uA$ part of the evolution equation for the angular momentum aspect produces an equation similar to that of Eq.~\eqref{eq:bondimassevo},
\begin{align}
\label{eq:bondiangmomevo}
\dot{N}_{A}&=D_{A}m+\frac{1}{4}D_{B}D_{A}D_{C}C^{BC}-\frac{1}{4}D^2D^{B}C_{AB}\nonumber\\
&\phantom{=.}+\frac{1}{4}D_{B}(C_{AC}N^{BC})+\frac{1}{2}C_{AC}D_{B}N^{BC}.
\end{align}
However, the terms in this equation cannot as clearly be classified as ``memory-like,'' ``ordinary-like,'' and ``null-like,'' analogous to those appearing in Eq.~\eqref{eq:bondimassevo} or~\eqref{eq:bondimassevo_null}. Therefore, before we compute the memory, we must first rewrite Eq.~\eqref{eq:bondiangmomevo} in terms of the function $\widehat{N}_{A}$, which can be thought of as an angular momentum that corresponds to the conserved super-Lorentz charges. We henceforth call $\widehat{N}_{A}$ the angular momentum aspect rather than $N_{A}$. According to Flanagan and Nichols's~\cite{Flanagan_2017} Eq.~(3.11), \(\widehat{N}_{A}\) is
\begin{align}
\label{eq:bondimomentumaspect}
\widehat{N}_{A}&\equiv N_{A}-u D_{A}m\nonumber\\
&\phantom{=.}-\frac{1}{16}D_{A}(C_{BC}C^{BC})-\frac{1}{4}C_{AB}D_{C}C^{BC}.
\end{align}
Using Eq.~\eqref{eq:bondiangmomevo} in the retarded time derivative of Eq.~\eqref{eq:bondimomentumaspect} produces the result
\begin{align}
\label{eq:angmomderiv}
\partial_{u}\widehat{N}_{A}&=\frac{1}{4}(D_{B}D_{A}D_{C}C^{BC}-D^2D^{B}C_{AB})\nonumber\\
&\phantom{=.}+\frac{1}{4}D_{B}(C_{AC}N^{BC})+\frac{1}{2}C_{AC}D_{B}N^{BC}\nonumber\\
&\phantom{=.}-\frac{1}{8}D_{A}(C_{BC}N^{BC})-\frac{1}{4}N_{AB}D_{C}C^{BC}\nonumber\\
&\phantom{=.}-\frac{1}{4}C_{AB}D_{C}N^{BC}-uD_{A}\dot{m}\nonumber\\
&=\frac{1}{4}(D_{B}D_{A}D_{C}C^{BC}-D^2D^{B}C_{AB})\nonumber\\
&\phantom{=.}-\Big[(\frac{3}{8}N_{AB}D_{C}C^{BC}-\frac{3}{8}C_{AB}D_{C}N^{BC})\nonumber\\
&\phantom{=.}-(\frac{1}{8}N^{BC}D_{B}C_{AC}-\frac{1}{8}C^{BC}D_{B}N_{AC})\Big]\nonumber\\
&\phantom{=.}-uD_{A}\dot{m}.
\end{align}
For the second equality, we have used
\begin{align}
N^{BC}D_{A}C_{BC}&=N^{BC}D_{B}C_{AC}+N_{AB}D_{C}C^{BC},\\
C^{BC}D_{A}N_{BC}&=C^{BC}D_{B}N_{AC}+C_{AB}D_{C}N^{BC}.
\end{align}
Finally, using the angular momentum aspect, we may write the evolution equation~\eqref{eq:bondiangmomevo} as
\begin{align}
\label{eq:bondiangmomevo_null}
\frac{1}{4}&(D_{B}D_{A}D_{C}C^{BC}-D^2D^{B}C_{AB})\nonumber\\
&\phantom{=.}=\partial_{u}(\widehat{N}_{A}+8\pi\mathcal{J}_{A})+uD_{A}\dot{m},
\end{align}
where
\begin{align}
\label{eq:radiatedangmom}
\dot{\mathcal{J}}_{A}&\equiv\frac{1}{64\pi}\Big[(3N_{AB}D_{C}C^{BC}-3C_{AB}D_{C}N^{BC})\nonumber\\
&\phantom{=.\frac{1}{64\pi}\big[}-(N^{BC}D_{B}C_{AC}-C^{BC}D_{B}N_{AC})\Big]
\end{align}
is the retarded time derivative of the angular-momentum radiated per unit solid angle. Akin to Eq.~\eqref{eq:bondimassevo_null}, we have written Eq.~\eqref{eq:bondiangmomevo_null} so that the first, second, and third terms on the right-hand side of the equation correspond to the memory that can be found in the shear as well as the ordinary and null memory contributions. As we will show next, Eq.~\eqref{eq:bondimassevo_null} produces the displacement memory while its counterpart, Eq.~\eqref{eq:bondiangmomevo_null}, produces the recently discovered spin memory. While we do not present an explicit equation for the CM memory effect, Eq.~\eqref{eq:bondimassevo_null} can be shown to contain terms that relate to the CM memory (see Sec.~\ref{sec:cmmemorycomputation} for more explanation).

\subsection{Computation of Memory}
\label{sec:memorycomputation}
Consider a spacetime in which the flux of energy and angular momentum to future null infinity vanishes before some early retarded time $u_{1}$, so that the news tensor $N_{AB}$ and the stress-energy tensor vanish there as well. Further, assume that sometime thereafter there is emission of gravitational waves, and that these fluxes again vanish for times after some $u_{2}>u_{1}$. The displacement memory is the effect that a pair of freely falling, initially comoving observers will then be able to observe a nonzero change in their relative position. This change is determined by changes to the spacetime of order $1/r$ and is given by a function known as the \emph{memory tensor},
\begin{align}
\Delta C_{AB}&\equiv C_{AB}(u_{2})-C_{AB}(u_1).
\end{align}
Here, we use the notation $\Delta f\equiv f(u_{2})-f(u_{1})$ where $f$ is some function of Bondi time.

We now write the memory tensor as the sum of an electric and a magnetic component. Motivated by how one may write a vector field on the two-sphere as the sum of a gradient (``electric'') and a curl (``magnetic'')\footnote{i.e., $V_{A}=D_{A}\Phi+\epsilon_{AB}D^{B}\Psi$.}, we have
\begin{align}
\label{eq:emdecomposition}
\Delta C_{AB}&=(D_{A}D_{B}-\frac{1}{2}q_{AB}D^2)\Delta\Phi+\epsilon_{C(A}D_{B)}D^{C}\Delta\Psi,
\end{align}
where $\Delta\Phi\equiv\Phi(u_2)-\Phi(u_1)$ and $\Delta\Psi\equiv\Psi(u_2)-\Psi(u_1)$ are scalar functions that represent the electric and magnetic components of the displacement memory and \(\epsilon_{AB}\) is just the Levi-Civita tensor on the two-sphere.

Because our Cauchy-characteristic extraction extracts the strain $h$, we now rewrite the BMS flux-balance laws, i.e., Eqs.~\eqref{eq:bondimassevo_null} and~\eqref{eq:bondiangmomevo_null}, in terms of this observable. Using the complex dyad introduced previously in Sec.~\ref{sec:conventions}, we construct the strain as a spin-weight $-2$ quantity:
\begin{align}
h\equiv\frac{1}{2}\bar{q}^{A}\bar{q}^{B}C_{AB}=\sum\limits_{\ell\geq2}\sum\limits_{|m|\leq \ell} h_{\ell m}\,\phantom{}_{-2}Y_{\ell m}(\theta,\phi).
\end{align}
Here we are only considering the \(1/r\) part of the strain. Generally the strain is computed using the full metric at asymptotic infinity---namely, $h\equiv\frac{1}{2}\bar{q}^{A}\bar{q}^{B}\gamma_{AB}$. However, the $1/r$ part of the strain is the only observable component at future null infinity and thus all we need to consider.

We now use Eqs.~\eqref{eq:bondimassevo_null} and~\eqref{eq:bondiangmomevo_null} to compute the memory $\Delta J$. But, to simplify this work we first write the memory in terms of its electric and magnetic components, i.e., $\Delta J=\Delta J^{(E)}+\Delta J^{(B)}$, where
\begin{subequations}
\begin{align}
\label{eq:ememorydef}
\Delta J^{(E)}&\equiv\frac{1}{2}\bar{q}^{A}\bar{q}^{B}\Delta C_{AB}^{(E)}(\Delta\Phi)\nonumber\\
&=\frac{1}{2}\bar{q}^{A}\bar{q}^{B}\Big[(D_{A}D_{B}-\frac{1}{2}q_{AB}D^2)\Delta\Phi\Big]\nonumber\\
&=+\frac{1}{2}\bar{\eth}^{2}\Delta\Phi,\\
\label{eq:bmemorydef}
\Delta J^{(B)}&\equiv\frac{1}{2}\bar{q}^{A}\bar{q}^{B}\Delta C_{AB}^{(B)}(\Delta\Psi)\nonumber\\
&=\frac{1}{2}\bar{q}^{A}\bar{q}^{B}\Big[\epsilon_{C(A}D_{B)}D^{C}\Delta\Psi\Big]\nonumber\\
&=-\frac{1}{2}i\bar{\eth}^{2}\Delta\Psi.
\end{align}
We reserve the letter $J$ to represent observables that we calculate using functions extracted from our simulations, such as the strain $h$, the news $\dot{h}$, or the Weyl scalars.
\end{subequations}
\subsubsection{Electric Memory}
\label{sec:ememorycomputation}
The electric component of the memory is the piece that arises from the scalar function $\Delta\Phi$. Using Eq.~\eqref{eq:emdecomposition}, the memory term in Eq.~\eqref{eq:bondimassevo_null} becomes
\begin{align}
\label{eq:memorycontraction}
\frac{1}{4}D^{A}D^{B}\Delta C_{AB}&=\frac{1}{8}(D^4-2D^{A}\left[D_{A},D_{B}\right]D^{B})\Delta\Phi\nonumber\\
&=\frac{1}{8}(D^4+2D^{A}q_{AB}D^{B})\Delta\Phi\nonumber\\
&=\frac{1}{8}D^2(D^2+2)\Delta\Phi\nonumber\\
&=\mathfrak{D}\Delta\Phi,
\end{align}
where
\begin{align}
\label{eq:operator}
\mathfrak{D}\equiv\frac{1}{8}D^2(D^2+2).
\end{align}
In computing Eq.~\eqref{eq:memorycontraction} we have used the fact that $[D_{A},D_{B}]D^{B}=-q_{AB}D^{B}$ on the two-sphere and used symmetry/antisymmetry to remove the dependence on the magnetic term $\Delta\Psi$. We act on Eq.~\eqref{eq:memorycontraction} with $\mathfrak{D}^{-1}$ to obtain an expression for $\Delta\Phi$. But, because $\mathfrak{D}$ maps the $\ell=0,1$ modes to zero, $\mathfrak{D}^{-1}$'s action on these modes is ambiguous. Therefore, to avoid this complication we construct $\mathfrak{D}^{-1}$ so that it maps the $\ell=0,1$ modes to zero. Note that this choice has no effect on the strain since it is a spin-weight $-2$ function, and will thus be independent of these modes. By acting on Eq.~\eqref{eq:memorycontraction} with $\mathfrak{D}^{-1}$ and combining the result with the expression from Eq.~\eqref{eq:bondimassevo_null}, we then obtain
\begin{align}
\label{eq:ememory}
\Delta\Phi=\mathfrak{D}^{-1}\left[\Delta m+4\pi\left(\frac{1}{32\pi}\int_{u_{1}}^{u_{2}}N_{AB}N^{AB}\,du\right)\right].
\end{align}
Using
\begin{align}
C_{AB}&=\frac{1}{2}\big(q_{A}q_{B}h+\bar{q}_{A}\bar{q}_{B}\bar{h}\big),
\end{align}
which follows from the symmetric, trace-free condition of the shear tensor, we find that we may write Eq.~\eqref{eq:ememory} as
\begin{align} \label{eq:ememorystrainj}
\Delta\Phi=\mathfrak{D}^{-1}\left[\Delta m+4\pi\left(\frac{1}{16\pi}\int_{u_{1}}^{u_{2}}\dot{h}\dot{\bar{h}}\,du\right)\right].
\end{align}
Thus, the electric component of the memory can readily be found by combining the results of Eqs.~\eqref{eq:ememorydef} and~\eqref{eq:ememory},
\begin{align}
\label{eq:ememoryfinal}
\Delta J^{(E)}=\frac{1}{2}\bar{\eth}^{2}\mathfrak{D}^{-1}\left[\Delta m+\frac{1}{4}\int_{u_{1}}^{u_{2}}\dot{h}\dot{\bar{h}}\,du\right],
\end{align}
with the $\Delta m$ term as the ordinary contribution and the $\dot{h}\dot{\bar{h}}$ term as the null contribution. Equation~\eqref{eq:ememoryfinal} could also be written with $\eth^{-2}$ since this operator is equivalent to $\frac{1}{8}\bar{\eth}^{2}\mathfrak{D}^{-1}$ when acting on spin-weight 0 functions. But, we choose to use $\mathfrak{D}$ for numerical purposes. At this point, it remains to compute the Bondi mass aspect in terms of the strain and the Weyl scalar $\Psi_{2}$. As is shown in Appendix~\ref{a:bondimass}, by Eq.~\eqref{eq:bondimass}, the result one obtains is
\begin{align}
m=-\text{Re}\left[\Psi_{2}+\frac{1}{4}\dot{h}\bar{h}\right],
\end{align}
where $\text{Re}$ denotes the real part.

\subsubsection{Magnetic Memory}
\label{sec:bmemorycomputation}

To compute the magnetic memory, we use Eq.~\eqref{eq:bondiangmomevo_null} and proceed in a similar manner to the above calculation of the electric memory. By replacing $C_{AB}$ with $\Delta C_{AB}$, Eq.~\eqref{eq:bondiangmomevo_null} can be written as
\begin{align}
\label{eq:magmembeforecontractionone}
&\frac{1}{4}(D_{B}D_{A}D_{C}\Delta C^{BC}-D^2D^{B}\Delta C_{AB})\nonumber\\
&\phantom{=.}=\Delta\Big[\partial_{u}(\widehat{N}_{A}+8\pi\mathcal{J}_{A})+uD_{A}\dot{m}\Big].
\end{align}
Using Eq.~\eqref{eq:emdecomposition} in Eq.~\eqref{eq:magmembeforecontractionone} and making use of the identity $D_{A}[D^4,D^{A}]\Delta\Psi=D^{2}(2D^2+1)\Delta\Psi$, which follows from $D_{A}[D^{4},D^{B}]f(\theta,\phi)=D_{A}D^{B}(2D^2+1)f(\theta,\phi)$, we obtain
\begin{align}
\label{eq:bmemscalar}
\frac{1}{4}(D_{B}D_{A}D_{C}\Delta C^{BC}-D^2D^{B}\Delta C_{AB})=\epsilon_{AC}D^{C}\mathfrak{D}\Delta\Psi,
\end{align}
Note that the electric component $\Delta\Phi$ vanishes because of various commutation relations similar to the one above. Therefore, we have the relation
\begin{align}
\label{eq:magmembeforecontractiontwo}
\epsilon_{AC}D^{C}\mathfrak{D}\Delta\Psi=\Delta\left[\partial_{u}(\widehat{N}_{A}+8\pi\dot{\mathcal{J}}_{A})+uD_{A}\dot{m}\right].
\end{align}
If we now contract Eq.~\eqref{eq:magmembeforecontractiontwo} with the function $\epsilon^{AB}D_{B}$, since \(\epsilon^{AB}=\frac{1}{2}i(q^{A}\bar{q}^{B}-\bar{q}^{A}q^{B})\), we obtain
\begin{align}
\label{eq:psibeforeinverse}
\mathfrak{D}D^2\Delta\Psi&=\Delta\epsilon^{AB}D_{B}\left[\partial_{u}(\widehat{N}_{A}+8\pi\dot{\mathcal{J}}_{A})+uD_{A}\dot{m}\right]\nonumber\\
&=\Delta\text{Im}\left[\eth\partial_{u}(\widehat{\overline{N}}+8\pi\overline{\mathcal{J}})\right],
\end{align}
where $\text{Im}$ denotes the imaginary part and
\begin{align}
\widehat{N}\equiv q_{A}\widehat{N}^{A}\quad\text{and}\quad\mathcal{J}\equiv q_{A}\mathcal{J}^{A}.
\end{align}
Note that the Bondi mass aspect term drops out because of the commutativity of the covariant derivatives when acting on a scalar function and the antisymmetry of the Levi-Civita tensor. Consequently, by acting on Eq.~\eqref{eq:psibeforeinverse} with $\mathfrak{D}^{-1}D^{-2}$ and using Eq.~\eqref{eq:radiatedangmom} we have
\begin{subequations}
\begin{align}
\label{eq:psiequation}
\Delta\Psi&=\mathfrak{D}^{-1}D^{-2}\Delta\text{Im}\left[\eth\partial_{u}(\widehat{\overline{N}}+8\pi\overline{\mathcal{J}})\right]\\
&=\mathfrak{D}^{-1}D^{-2}\Delta\text{Im}\bigg\lbrace\eth(\partial_{u}\widehat{\overline{N}})+\frac{1}{8}\eth\bar{q}^{A}\nonumber\\
&\phantom{=.\Delta}\Big[(3N_{AB}D_{C}C^{BC}-3C_{AB}D_{C}N^{BC})\nonumber\\
&\phantom{=.\Delta\Big[}-(N^{BC}D_{B}C_{AC}-C^{BC}D_{B}N_{AC})\Big]\bigg\rbrace.
\end{align}
\end{subequations}
Expressing the angular momentum flux quantities on the right-hand side in terms of the observable $h$ gives
\begin{subequations}
	\begin{align}
	N_{AB}D_{C}C^{BC}&=\text{Re}\big[q_{A}\dot{h}\bar{\eth}\bar{h}\big],\\
	\label{eq:C_{AB}D^{C}N^{BC}}
	C_{AB}D_{C}N^{BC}&=\text{Re}\big[q_{A}h\bar{\eth}\dot{\bar{h}}\big],\\
	N^{BC}D_{B}C_{AC}&=\text{Re}\big[q_{A}\dot{\bar{h}}\bar{\eth}h\big],\\
	C^{BC}D_{B}N_{AC}&=\text{Re}\big[q_{A}\bar{h}\bar{\eth}\dot{h}\big],
	\end{align}
\end{subequations}
Thus, by combining everything together and using the result of Eq.~\eqref{eq:bmemorydef}, we find
\begin{align}
\label{eq:bmemoryfinal}
\Delta J^{(B)}&=\frac{1}{2}i\bar{\eth}^{2}\mathfrak{D}^{-1}D^{-2}\Delta\text{Im}\bigg\lbrace\bar{\eth}(\partial_{u}\widehat{N})\nonumber\\
&\phantom{=.}+\frac{1}{8}\left[\eth(3h\bar{\eth}\dot{\bar{h}}-3\dot{h}\bar{\eth}\bar{h}+\dot{\bar{h}}\bar{\eth}h-\bar{h}\bar{\eth}\dot{h})\right]\bigg\rbrace.
\end{align}

Next, we need the angular momentum aspect in terms of the strain and the Weyl scalar $\Psi_{1}$. As is shown in Appendix~\ref{a:bondiangmom}, by Eq.~\eqref{eq:imangmom}, the result one obtains is
\begin{align}
\text{Im}\Big[\bar{\eth}(\partial_{u}\widehat{N})\Big]=\text{Im}\left\lbrace2\bar{\eth}\dot{\Psi}_{1}-\frac{1}{4}\bar{\eth}\left[\partial_{u}(\bar{h}\eth h)\right]\right\rbrace.
\end{align}
As is illustrated by either Eq.~\eqref{eq:psiequation} or~\eqref{eq:bmemoryfinal}, the magnetic component of the memory is the total derivative with respect to retarded time of some scalar function, whereas the electric component of the memory contains terms that are either net changes, i.e., the $\Delta m$ term, or retarded time integrals, i.e., the $\dot{h}\dot{\bar{h}}$ term. Consequently, since the magnetic memory does not have such terms, one might presume that the magnetic memory vanishes, i.e., that the net change in the magnetic component of the strain is zero. Currently, this is unknown~\cite{Flanagan_2017, Winicour_2014, Bieri_2014, ashtekar2019compact}. But, it \emph{is} known that the retarded time integral of the magnetic memory does not vanish; this is what we refer to as the spin memory effect. We explore the conjectured vanishing feature of the magnetic memory in Sec.~\ref{sec:bmemory} and the spin memory in Sec.~\ref{sec:spinmemory}.

Equipped with both Eqs.~\eqref{eq:ememoryfinal} (the electric memory) and~\eqref{eq:bmemoryfinal} (the magnetic memory), we may now compute the electric and magnetic memory contributions to the strain by expressing each of these functions as a sum over spin-weighted spherical harmonics and acting with the inverse operators accordingly,
\begin{subequations}
\begin{align}
D^{-2}\phantom{}Y_{\ell m}&=[-\ell(\ell+1))]^{-1}Y_{\ell m},\\
\mathfrak{D}^{-1}Y_{\ell m}&=\left[\frac{1}{8}(\ell-1)\ell(\ell+1)(\ell+2)\right]^{-1}Y_{\ell m}.
\end{align}
\end{subequations}
We thus obtain the spin-weighted spherical harmonic representation of the memory
\begin{align}\label{eq:delta_J_modes}
\Delta J(\theta,\phi)=\sum\limits_{\ell\geq2}\sum\limits_{|m|\leq\ell}\Delta J_{\ell m}\,\phantom{}_{-2}Y_{\ell m}(\theta,\phi),
\end{align}
which we can use to compare the memory modes to those of the CCE extracted strain produced in our various numerical relativity simulations.

\subsubsection{CM Memory}
\label{sec:cmmemorycomputation}
Finally, we now illustrate how one can realize that Eq.~\eqref{eq:ememoryfinal} contains terms contributing to the CM memory. According to Eq.~\eqref{eq:magmembeforecontractiontwo}, we have
\begin{align}
\partial_{u}\Delta\widehat{N}_{A}=\frac{1}{8}\epsilon_{AC}D^{C}\mathfrak{D}\Delta\Psi-8\pi\Delta\dot{\mathcal{J}}_{A}-uD_{A}\Delta\dot{m}.
\end{align}
If we then contract this equation with \(D^{A}\) and take the real part of the entire equation, we obtain
\begin{align}
\partial_{u}\text{Re}(\bar{\eth}\widehat{N})&=-8\pi\text{Re}(\bar{\eth}\dot{\mathcal{J}})-uD^{2}\dot{m}\nonumber\\
&=-8\pi\text{Re}(\bar{\eth}\dot{\mathcal{J}})-\partial_{u}(uD^2m)+D^{2}m,
\end{align}
since the Bondi mass aspect term is a purely real quantity. By rearranging this equation and then entering the results back into the ordinary part of Eq.~\eqref{eq:ememoryfinal}, we obtain
\begin{align}
\label{eq:bondimasssuperLorentz}
\Delta J^{(E)}_{\text{ordinary}}&=\frac{1}{2}\bar{\eth}^{2}\mathfrak{D}^{-1}\Delta\bigg\lbrace(m+u\dot{m})+\nonumber\\
&\phantom{=.\frac{1}{2}\bar{\eth}\bar{\eth}\mathfrak{D}^{-1}\Delta\bigg\lbrace}\partial_{u}D^{-2}\text{Re}\left[\bar{\eth}(\widehat{N}+8\pi\mathcal{J})\right]\bigg\rbrace.
\end{align}
When written in this manner, it is now clear how the ordinary part of the electric memory can be realized as containing terms involving the retarded time derivative of the real part of the super-Lorentz charges, which are a part of the $\widehat{N}$ term, and the angular momentum flux. Even though this is somewhat trivial since we have simply changed the Bondi mass aspect by a function that is zero, Eq.~\eqref{eq:bondimasssuperLorentz} nonetheless illustrates how the ordinary part of the electric memory can be broken up into not only a displacement contribution (the first two terms), but also the time derivative of a CM contribution (the terms with the $\partial_{u}$ in front of them). To obtain the full expression for the CM memory, the remaining component that is needed is the null contribution, which can, in principle, be extracted from the energy flux. Joining this component with the ordinary CM memory contribution in Eq.~\eqref{eq:bondimasssuperLorentz} gives the full expression for the CM memory in terms of its ordinary and null parts. We explore the CM memory further with numerical results in Appendix~\ref{a:cmmem}.

\section{Results}
\label{sec:results}
We now compute the electric and magnetic components of the memory for various binary black hole simulations run using the code $\texttt{SpEC}$. Each of these merger simulations corresponds to an entry in the public SXS Catalog~\cite{Boyle_2019} and collectively encompasses both equal and unequal masses, spinning and nonspinning black holes, and configurations that are either precessing or nonprecessing. We provide the main parameters of these simulations in Table~\ref{mergers}.
\begin{table}[H]
	\caption{\label{mergers}%
		Primary parameters of the various BBH mergers analyzed in this paper. We use the mass and effective spin values that are obtained at the simulation's relaxation time~\cite{Boyle_2019}. While these are the runs that we show in this paper, many others have been used to understand and refine our conclusions. The spin vectors of 1389 are
		$\chi_{1}=(-0.2917,+0.2005,-0.3040)$ and $\chi_{2}=(-0.01394,+0.4187,+0.1556)$.}
	\centering
	\begin{ruledtabular}
		\begin{tabular}{ccccc}
			SXS:BBH: & Classification & $M_{1}/M_{2}$ & $\chi_{\text{eff}}$ & $N_{\text{orbits}}$ \\ [0.5ex] 
			\colrule
			1155 & Nonspinning & $1.000$ & $+2.617\times10^{-5}$ & $40.64$ \\ 
			0554 & Nonspinning & $2.000$ & $+4.879\times10^{-5}$ & $19.25$ \\ 
			1412 & Spinning & $1.630$ & $+1.338\times10^{-1}$ & $145.1$ \\
			1389 & Precessing & $1.633$ & $-1.293\times10^{-1}$ & $140.4$ \\
			0305 & GW150914 & $1.221$ & $-1.665\times10^{-2}$ & $15.17$ \\
		\end{tabular}
	\end{ruledtabular}
\end{table}
Each simulation produces a GW strain computed by Regge-Wheeler-Zerilli (RWZ) extraction at a series of spheres of finite radius and then extrapolates the strain to future null infinity~\cite{Boyle_2019}. This is the strain that can be found in the SXS Catalog. Like Pollney and Reisswig~\cite{Pollney_2011}, we find, however, that this method for constructing the strain does not seem to be able to resolve the memory. Consequently, we instead compute the strain using CCE.

Fortunately, each of our BBH simulations also produces the metric and its derivatives on a series of world tubes, where each world tube is a coordinate two-sphere dragged through time that provides the inner boundary conditions for the CCE module from the code $\texttt{SpECTRE}$~\cite{moxon2020improved,SpECTRE}. We use this CCE module to explicitly compute the strain $h$ at future null infinity. Note that we use the variable $h$ to represent the strain thus obtained from CCE, while the variable $J$ has been reserved for the strain computed from the BMS flux-balance laws. These should be identical in the absence of numerical error. Furthermore, unlike earlier implementations of CCE that exhibited the resolution of the strain $(2,0)$ mode~\cite{Pollney_2011}, the $\texttt{SpECTRE}$ CCE module computes the strain directly, like~\cite{Bishop_2013}. Consequently, there is no need to compute the news first and then integrate it with respect to retarded time, which could introduce errors from the choice of integration constants.

Within the SXS Catalog, most of the BBH simulations follow only a few tens of binary orbits. PN computations of memory, however, include effects that are obtained by integrating over the waveform starting at $u\rightarrow-\infty$. Accordingly, we hybridize the numerical strain obtained from CCE with a PN waveform corresponding to the same BBH merger (see Sec.~\ref{sec:cce}) using the python packages $\texttt{GWFrames}$ and $\texttt{Post-Newtonian}$~\cite{GWFrames,PostNewtonian}. When using \texttt{Post-Newtonian}, we also modified the code to include memory terms up to 3PN order. With this scheme, we find that we can resolve the traditional and most prominent $m=0$ memory modes, as well as other $m\not=0$ modes that exhibit both the displacement and spin memory effects.

Last, it should be noted that we primarily use the python package $\texttt{scri}$ to perform our analysis~\cite{scri_url, Boyle2013, Boyle2016, Boyle2014}.

\subsection{CCE vs Extrapolation}
\label{sec:ccevsextrapolation}

\begin{figure*}
	\label{fig:ExtComparison}
	\centering
	\includegraphics[width=0.329\textwidth]{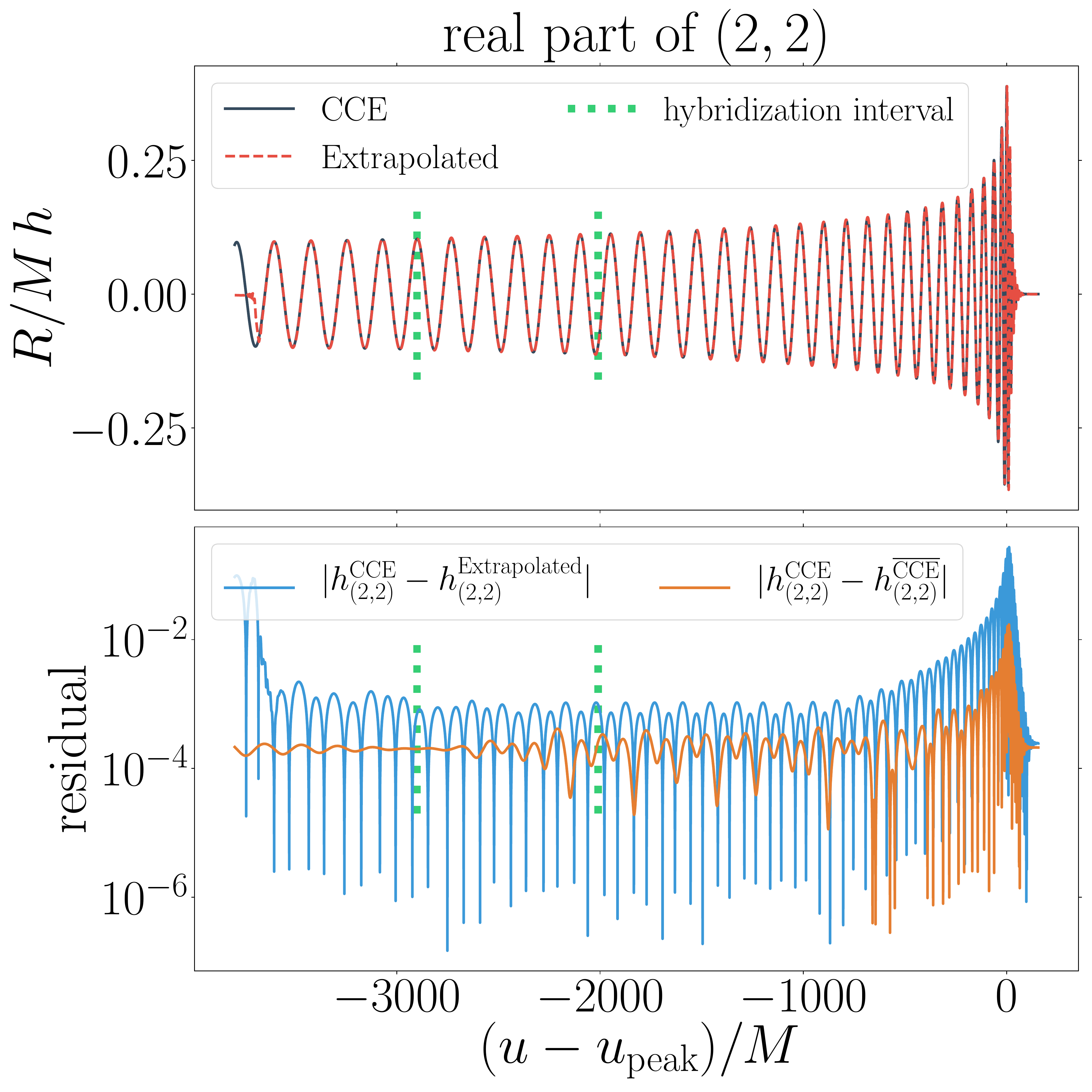}
	\includegraphics[width=0.329\textwidth]{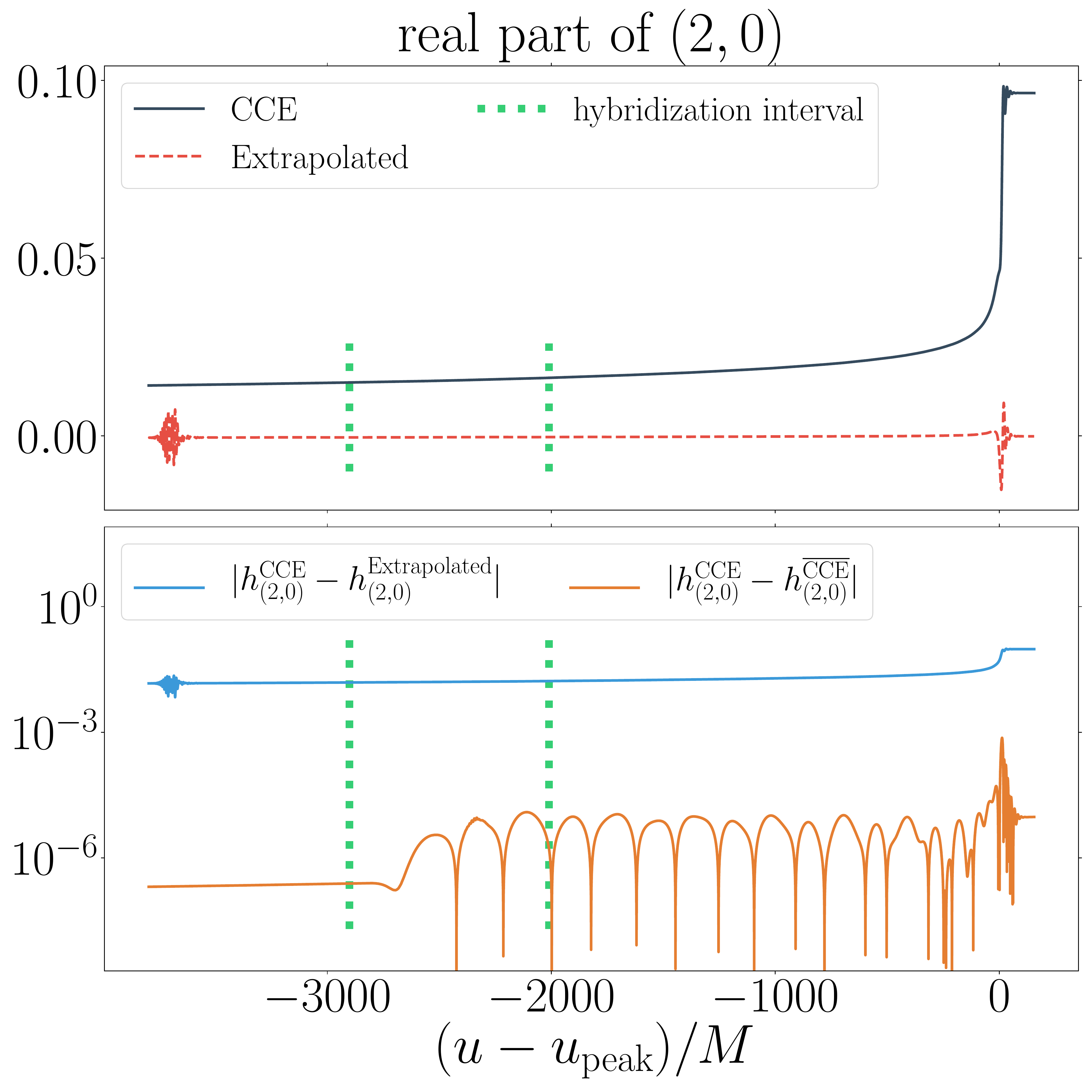}
	\includegraphics[width=0.329\textwidth]{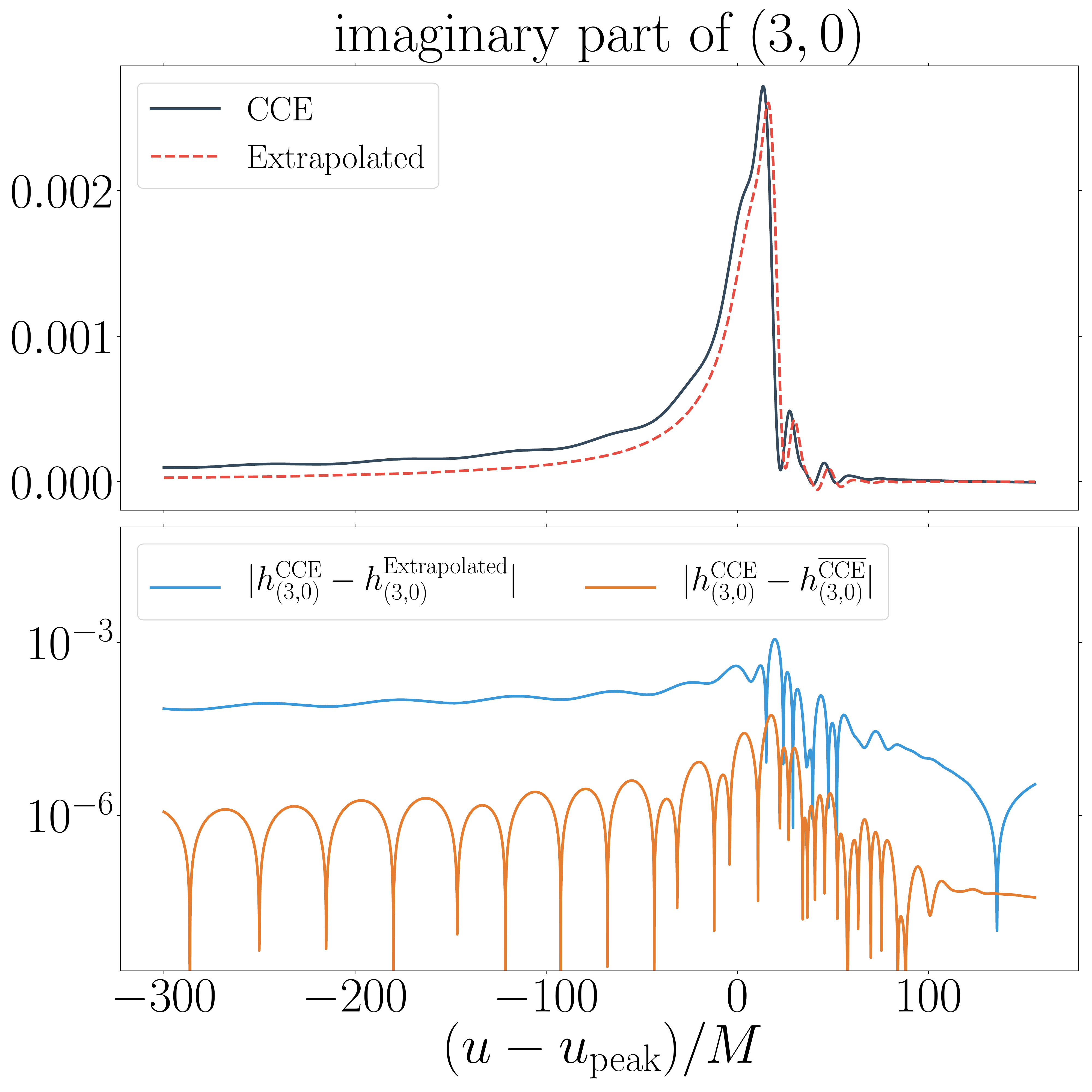}
	\caption{Comparison of the strain computed by CCE versus RWZ extraction followed by extrapolation to future null infinity, for several spin-weight $-2$ spherical harmonic modes of the SXS simulation SXS:BBH:0305. On each plot, we show the interval over which the hybridization between CCE and PN is performed, i.e., before this interval the waveform is purely from a post-Netwonian calculation while after this interval the waveform is purely from numerical computations. In the bottom row of each plot we provide the residuals and an estimate of the error in the CCE waveform, $|h_{(\ell,m)}^{\text{CCE}}-h_{(\ell,m)}^{\overline{\text{CCE}}}|$, where $h_{(\ell,m)}^{\text{CCE}}$ is the highest resolution waveform of SXS:BBH:0305 and $h_{(\ell,m)}^{\overline{\text{CCE}}}$ is the next highest resolution waveform for the same binary system. We align the waveforms in both time and phase around $u_{\text{peak}}$, which is where the $L^{2}$ norm of the strain achieves its maximum. See Table~\ref{mergers} for the parameters of SXS:BBH:0305.}
\end{figure*}

We first compare the strain that we compute using two distinct extraction methods: (1) RWZ extraction followed by extrapolation to future null infinity and (2) CCE plus a PN hybridization. In Fig.~\ref{fig:ExtComparison}. we compare three different spin-weight $-2$ spherical harmonic modes of the strain for the numerical simulation SXS:BBH:0305, which is a simulation of GW150914 (see Table~\ref{mergers}). We compare the $(2,2)$, $(2,0)$, and $(3,0)$ modes from CCE/PN hybrids to those from extrapolated RWZ waveforms. Each one of these modes corresponds to the most prominent mode for the strain as well as the electric and magnetic memory (see Secs.~\ref{sec:ememory} and~\ref{sec:bmemory}). We also show an estimate of the error in the CCE waveform $|h_{(\ell,m)}^{\text{CCE}}-h_{(\ell,m)}^{\overline{\text{CCE}}}|$, where $h_{(\ell,m)}^{\text{CCE}}$ is the highest resolution waveform available for SXS:BBH:0305 and $h_{(\ell,m)}^{\overline{\text{CCE}}}$ is the next highest resolution. While there is also some numerical error that comes from the Cauchy-characteristic extraction, we find that these errors are of order $10^{-10}$ and thus negligible in comparison to the Cauchy evolution's resolution error. Consequently, for all the plots in this paper, we only present the error that comes from the Cauchy evolution.

As can be seen in the plots in Fig~\ref{fig:ExtComparison}, the CCE and extrapolated waveforms coincide well for the $(2,2)$ mode. However, the extrapolation method fails to capture the memory contribution to the $(2,0)$ electric memory mode, but does recover the quasinormal mode ringdown near the peak of the waveform. Curiously, the extrapolated waveform \emph{does} contain nontrivial contributions to the imaginary part of the $(3,0)$ magnetic memory mode, but does not determine the time integral of the mode, which is the main contribution to the spin memory, accurately ($\roughly$50\% of the values seen in CCE for the runs we studied). Thus, the importance of using CCE can readily be seen: while previous extrapolation-based extraction schemes cannot accurately resolve memory effects, the current implementation of $\texttt{SpECTRE}$'s CCE can.

\subsection{Checking the Flux-balance Laws}
\label{sec:fluxbalance}
As shown in Sec.~\ref{sec:memorycomputation}, using Eqs.~\eqref{eq:ememoryfinal} and~\eqref{eq:bmemoryfinal}, one can compute the memory \(\Delta J(\theta,\phi)\), which is the change in the strain between the retarded times corresponding to the nonradiative\footnote{A BBH coalescence is never truly nonradiative at future infinity; here we assume that future infinity is approximately nonradiative at both early and late retarded times.} regimes that exist before and after the passage of radiation. However, the flux-balance laws---Eqs.~\eqref{eq:bondimassevo_null} and~\eqref{eq:bondiangmomevo_null}---from which the memory effects are computed should be true for any given retarded time. This version of these BMS flux-balance laws is called the finite time version, rather than the global version. Thus, to see if our Cauchy-characteristic extraction is performing as we expect it to for the strain as well as the Weyl scalars $\Psi_{1}$ and $\Psi_{2}$, we can compare the strain $h$ as obtained from CCE to the ``flux-balance strain,''
\begin{align}
J&\equiv\sum\limits_{\ell \geq2}\sum\limits_{|m|\leq\ell}J_{\ell m}\,\phantom{}_{-2}Y_{\ell m}(\theta,\phi)\nonumber\\
&=\sum\limits_{\ell \geq2}\sum\limits_{|m|\leq\ell}(J_{\ell m}^{(E)}+J_{\ell m}^{(B)})\,\phantom{}_{-2}Y_{\ell m}(\theta,\phi),
\end{align}
where \(J_{\ell m}^{(E)}\) and \(J_{\ell m}^{(B)}\) take on the same functional form as the spin-weighted spherical harmonic decompositions of \(\Delta J^{(E)}\) and \(\Delta J^{(B)}\) coming from Eqs.~\eqref{eq:ememoryfinal} and~\eqref{eq:bmemoryfinal}, but are now also functions of the retarded time $u$, i.e., the operator $\Delta$ from Eqs.~\eqref{eq:ememoryfinal} and~\eqref{eq:bmemoryfinal} is removed. Put differently, we wish to check the consistency of
\begin{align}
h=J^{(E)}+J^{(B)}
\end{align}
up to the error of the corresponding Cauchy evolution.\\
\begin{figure*}
	\label{fig:BondiCheck}
	\centering
	\includegraphics[width=0.329\linewidth]{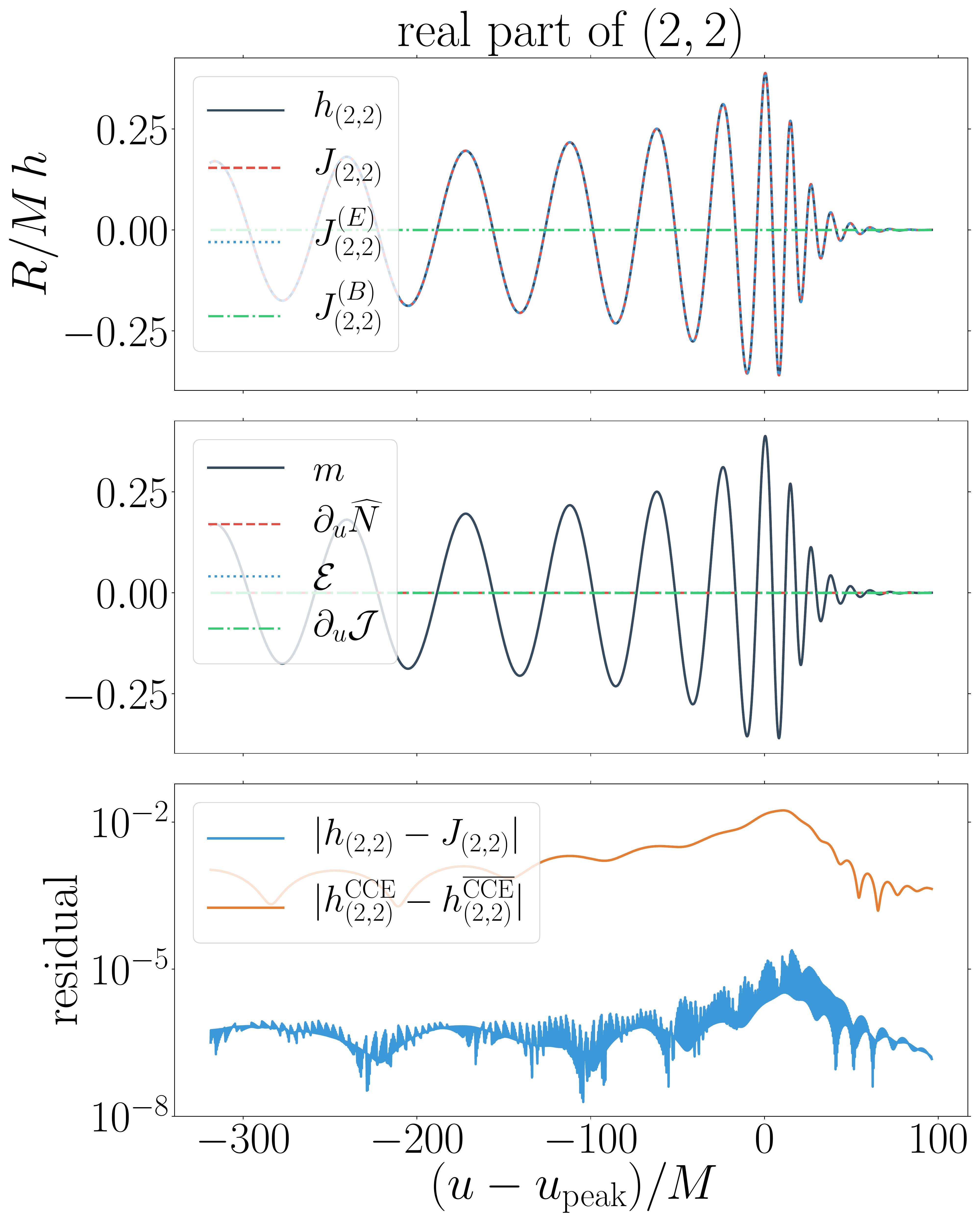}
	\includegraphics[width=0.329\linewidth]{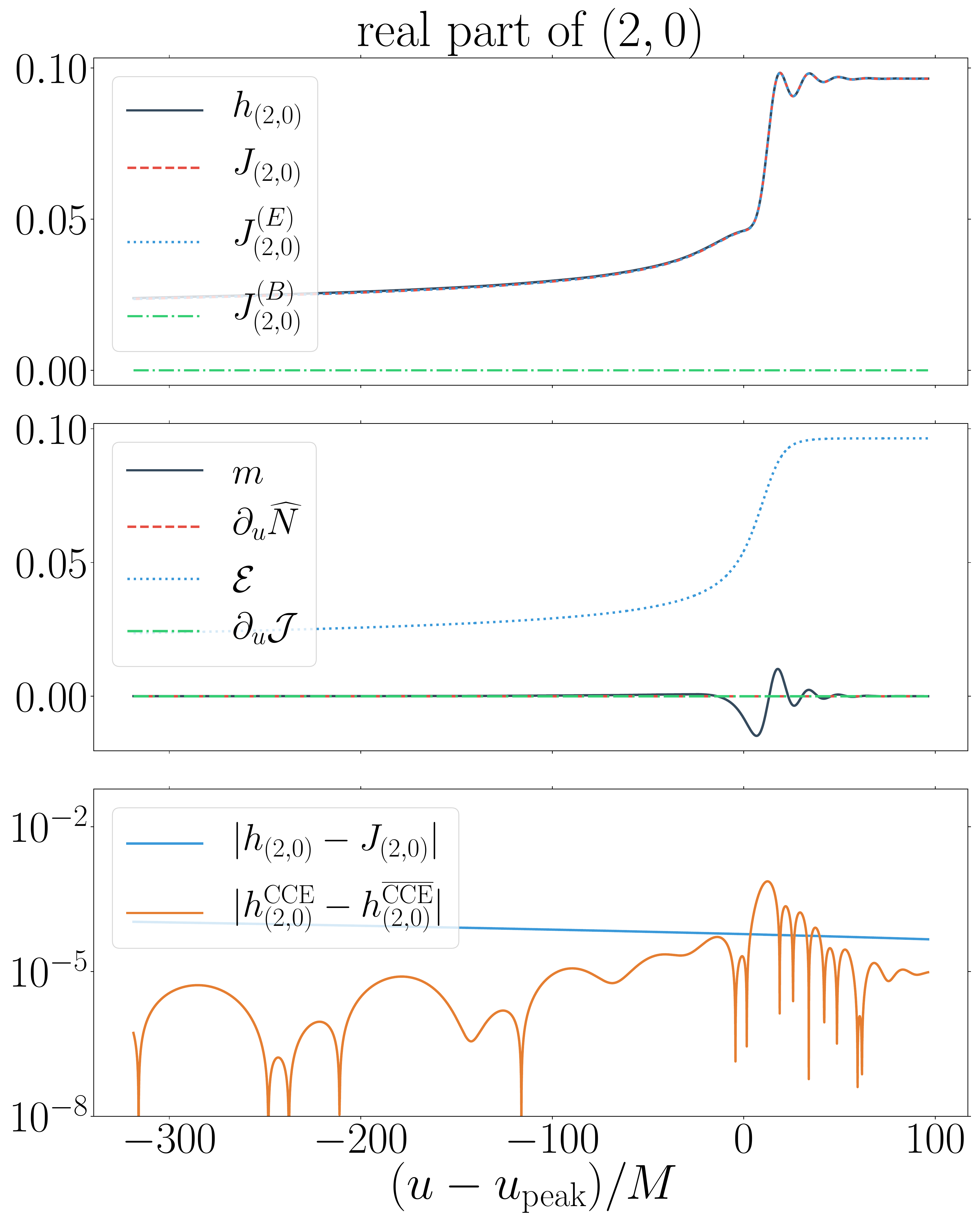}
	\includegraphics[width=0.329\linewidth]{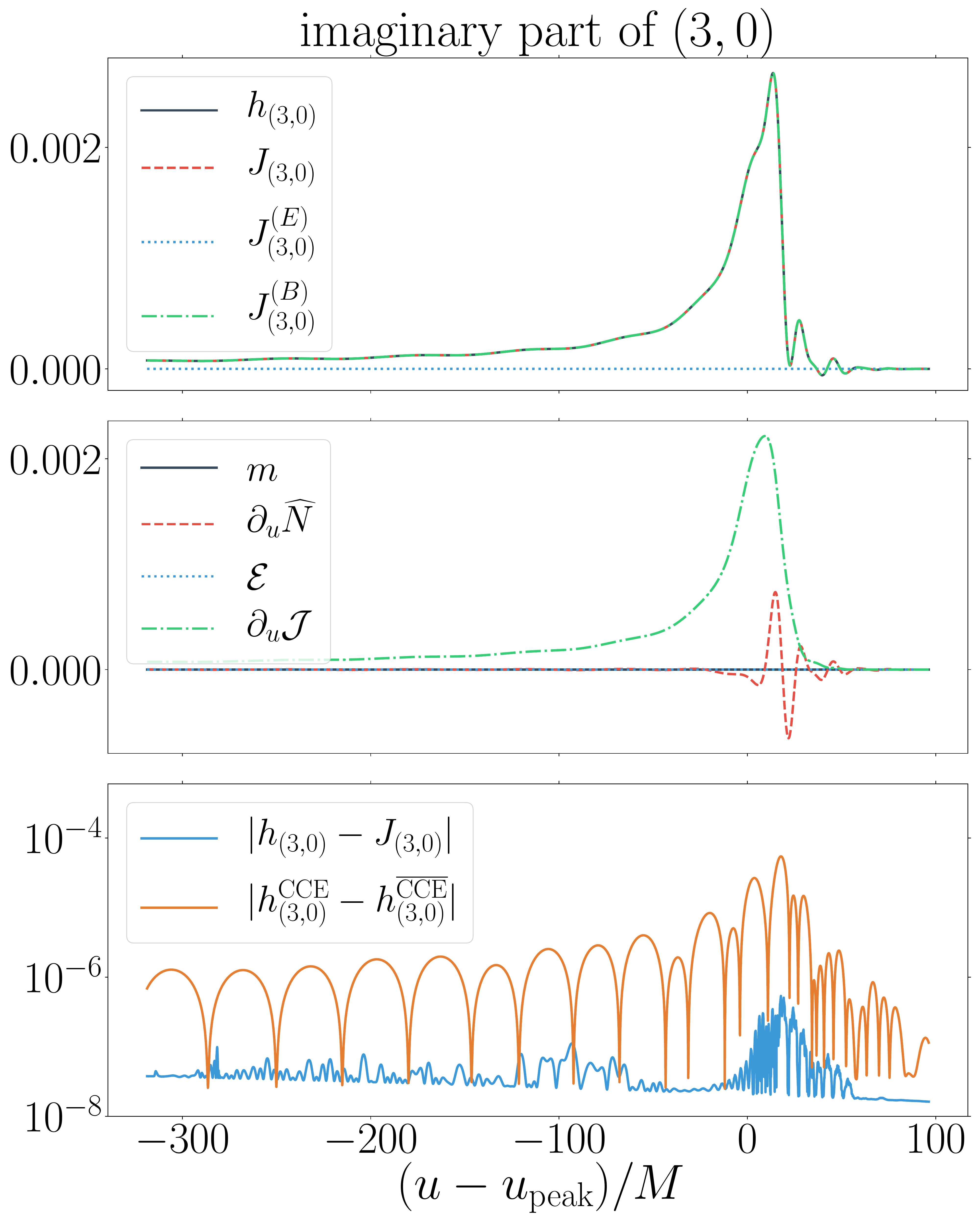}
	\caption{Comparison of the strain extracted using CCE from SXS simulation SXS:BBH:0305 versus the strain computed using the BMS flux-balance laws, Eqs.~\eqref{eq:ememoryfinal} and~\eqref{eq:bmemoryfinal}, without the $\Delta$ operator. Each column shows a particular spin-weight $-2$ mode. The top row shows the extracted strain (black/solid), the strain computed from the BMS flux-balance laws (red/dashed), and its corresponding electric (blue/dotted) and magnetic (green/dashed/dotted) components coming from Eqs.~\eqref{eq:ememoryfinal} and~\eqref{eq:bmemoryfinal}. The middle row shows the contributions from the mass aspect (black/solid), the angular momentum aspect (red/dashed), the energy flux (blue/dotted), and the angular momentum flux (green/dashed/dotted). Because the energy flux contribution to the electric component of the strain is an integral over retarded time, we are free to change the electric component by a constant. We take this angle-dependent constant to be the final value of the extracted strain.}
\end{figure*}

In Fig.~\ref{fig:BondiCheck}, we compare the strain obtained from CCE to the strain computed from the BMS charges and fluxes. As in the comparison shown in Fig~\ref{fig:ExtComparison}, we show results for the $(2,2)$, $(2,0)$ and $(3,0)$ modes for the same NR simulation as before: SXS:BBH:0305. As can be seen, the two coincide with each other rather well, with the $(2,2)$ mode being the best, followed by the $(2,0)$ mode, and then the $(3,0)$ mode. Most important, though, one can observe through the $(2,0)$ and $(3,0)$ modes that the memory primarily comes from the null contribution, while the ordinary contribution appears to only capture the quasinormal mode behavior. Nevertheless, this is perhaps as expected because the majority of the simulations in the SXS Catalog experience little to no supertranslations or super-Lorentz transformations~\cite{Boyle_2016}. Consequently, there will be little to no BMS charges for the radiation to carry to future null infinity, which will make the contribution from the ordinary memory small compared to that of the null memory, i.e.,
\begin{subequations}
\begin{align}
\label{eq:approxemem}
\Delta J^{(E)}&\approx+\frac{1}{8}\bar{\eth}^{2}\mathfrak{D}^{-1}\left[\int_{u_{1}}^{u_{2}}\dot{h}\dot{\bar{h}}\,du\right],\\
\label{eq:approxbmem}
\Delta J^{(B)}&\approx+\frac{1}{16}i\bar{\eth}^{2}\mathfrak{D}^{-1}D^{-2}\Delta\text{Im}\nonumber\\
&\phantom{=.-1}\left[\eth(3h\bar{\eth}\dot{\bar{h}}-3\dot{h}\bar{\eth}\bar{h}+\dot{\bar{h}}\bar{\eth}h-\bar{h}\bar{\eth}\dot{h})\right].
\end{align}
\end{subequations}

In this work, our primary objective is to provide a statement on the measurability of the memory rather than any other phenomenon, such as quasinormal modes. Thus, we need to consider the function that represents the instantaneous memory effect as a function of time. As can be seen in Fig.~\ref{fig:BondiCheck} and as was just discussed, the observable that serves as a reasonable proxy for this is the null contribution to the flux-balance strain. Therefore, in the following sections, we will only examine the null contribution to the flux-balance strain and henceforth refer to this contribution as the system's overall memory. We represent the memory as a function of time as
\begin{align}
\Delta J(u)\equiv\sum\limits_{\ell\geq2}\sum\limits_{|m|\leq\ell}\Delta J_{\ell m}(u)\,\phantom{}_{-2}Y_{\ell m}(\theta,\phi).
\end{align}

From an observational standpoint, a GW observatory will only be able to measure the complete memory mode, i.e., a superposition of memory and quasinormal modes. Thus, to measure the memory effect, one needs to be able to filter the quasinormal mode frequencies so that only the frequencies corresponding to the memory remain. As we thoroughly explore in Sec.~\ref{sec:snrs}, performing such a postprocessing analysis of LIGO observations should indeed be feasible, thereby allowing for the measurement of the memory induced by a GW within an interferometer. As a result, since the null memory contribution contains no quasinormal mode contribution, this is a fair proxy for what LIGO would see once the quasinormal modes have been filtered out of the strain memory modes.

Note that we are free to change the null contributions to the electric and spin memories by constants, since they depend on certain energy and angular momentum fluxes that are computed by performing retarded time integrals. The need for these angle-dependent constants is a result of not knowing the past history of the numerical waveforms. Unless stated otherwise, we choose these constants so that the flux-balance strain has the same initial value as the CCE/PN hybrid strain.

\subsection{Electric Memory Modes}
\label{sec:ememory}
We now analyze the main memory modes obtained from numerical relativity by comparing them to PN theory and $\Delta J(u)$ via the functional forms of Eqs.~\eqref{eq:approxemem} and~\eqref{eq:approxbmem}, i.e., Eqs.~\eqref{eq:ememoryfinal} and~\eqref{eq:bmemoryfinal} but without the contribution coming from the negligible ordinary memory.\footnote{While the ordinary contribution to the strain is not negligible, seeing as it contains information about the quasinormal modes, the memory part of this contribution can indeed be considered to be negligible, as we argued through the results shown in Fig.~\ref{fig:BondiCheck}.} According to Favata~\cite{Favata_2009, Favata_2009_PN, Favata_2010}, the bulk of the electric memory should be in the real component of the nonoscillatory $(2,0)$ mode, with other contributions primarily persisting in the other $\ell=\text{even},\,m=0$ modes. But, as was also noted by Favata, there may be memory contributions from $m\not=0$ oscillatory modes, e.g., the $(3,\pm1)$ modes. Consequently, we examine results for not only the usual $m=0$ memory modes, but also a few of the potential $m\not=0$ oscillatory memory modes. We begin by first illustrating the agreement between our $(2,0)$ mode and what is expected according to PN theory.

\begin{figure}
	\label{fig:PNComparison_L2_M0}
	\centering
	\includegraphics[width=\columnwidth]{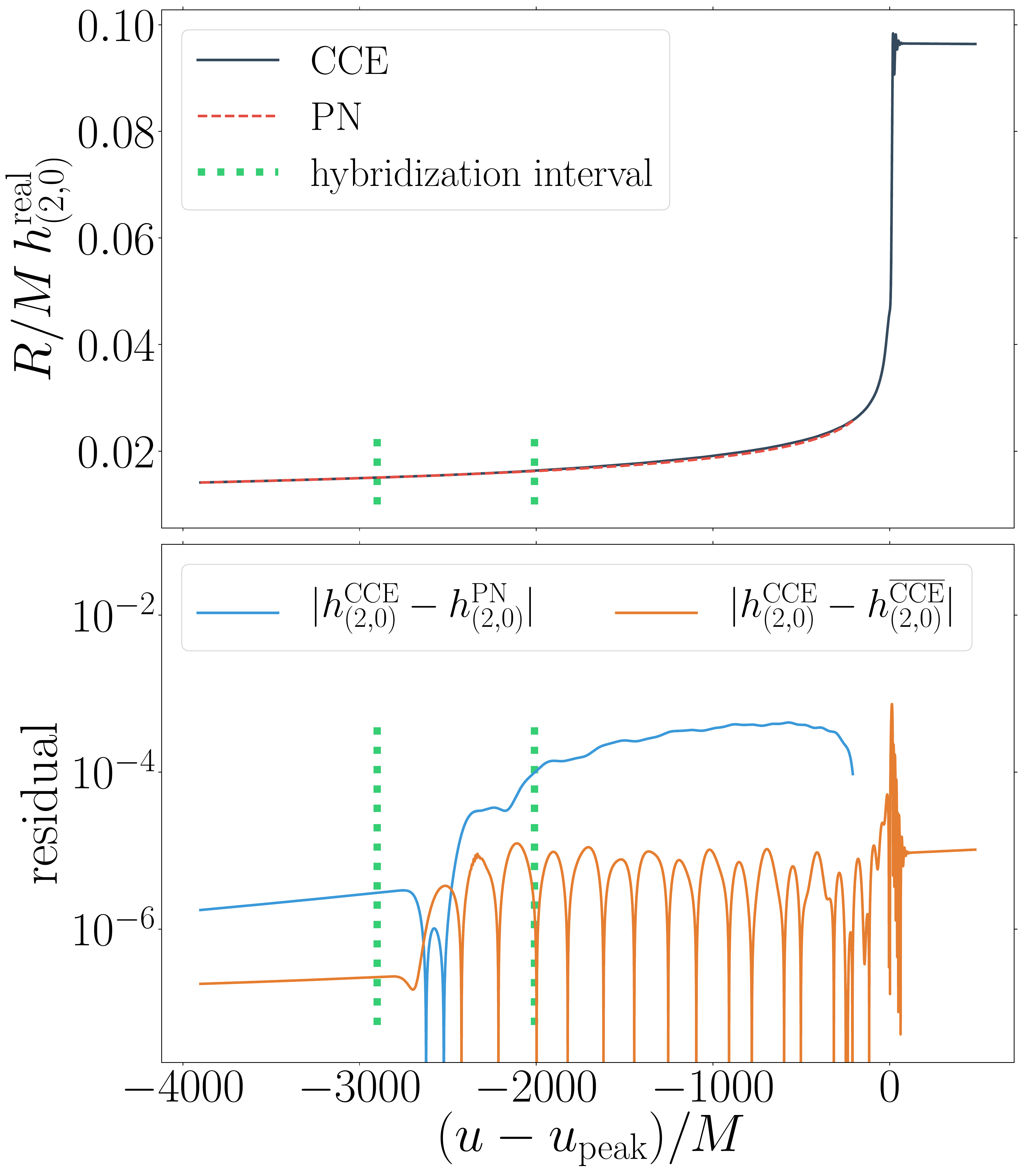}
	\caption{Comparison between the $(2,0)$ mode obtained from numerical relativity to that which is computed using PN theory. For reference, in the bottom plot we provide an estimate of the error in the CCE waveform, $|h_{(2,0)}^{\text{CCE}}-h_{(2,0)}^{\overline{\text{CCE}}}|$, where $h_{(2,0)}^{\text{CCE}}$ refers to the highest resolution waveform of SXS:BBH:0305 and $h_{(2,0)}^{\overline{\text{CCE}}}$ refers to the next highest resolution. The reason why the hybrid and the PN waveform are not identical before the hybridization interval is because there is numerical error that is introduced when aligning the two waveforms.}
\end{figure}

For this PN comparison, we consider SXS:BBH:0305. As in Fig.~\ref{fig:ExtComparison}, in Fig.~\ref{fig:PNComparison_L2_M0}, we show the agreement between CCE and PN in the top plot and provide a rough estimate of the numerical error in the bottom plot. As expected, the numerical waveform and the PN waveform coincide well during the inspiral, but then diverge from one another as the binary system approaches the merger phase.

\begin{figure}
	\label{fig:EMode_1155_M0}
	\centering
	\includegraphics[width=\columnwidth]{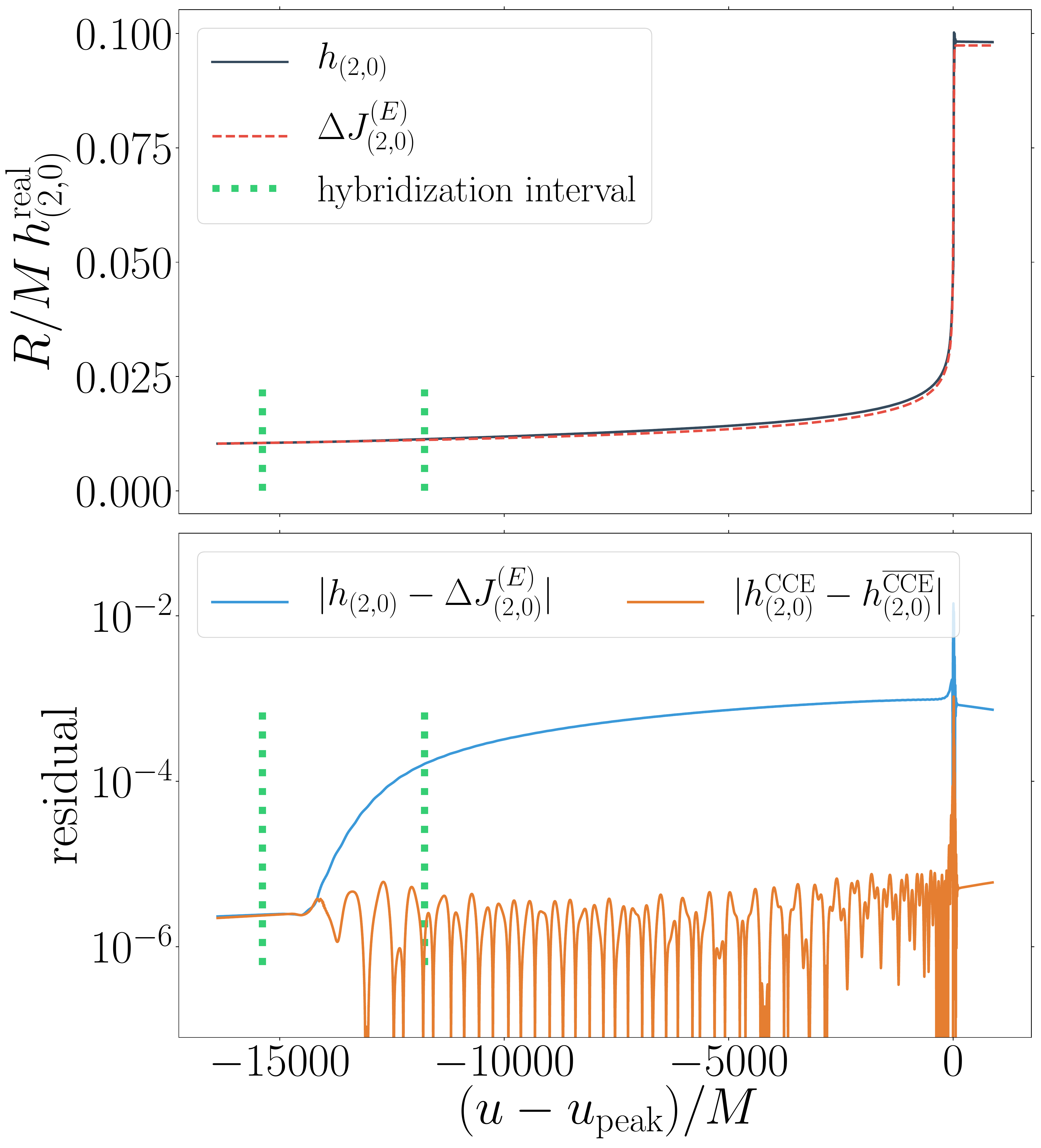}
	\caption{Comparison of the $(2,0)$ mode with the memory for an equal mass, nonspinning system (SXS:BBH:1155, see Table~\ref{mergers}).}
\end{figure}

Next, to illustrate the variation of the memory across various BBH parameters, we examine an equal mass and nonspinning system: SXS:BBH:1155. We again find that the main memory modes are the $m=0$ modes, with both of the $(2,0)$ and $(4,0)$ modes taking on values that are larger than the corresponding numerical error. However, the other $m=0$ modes acquire values that are smaller than can be resolved at this run's numerical resolution. Moreover, we find that both of the $(2,0)$ and $(4,0)$ modes coincide rather well with the instantaneous memory from Eqs.~\eqref{eq:approxemem} and~\eqref{eq:approxbmem}, as illustrated in Fig.~\ref{fig:EMode_1155_M0}.

\begin{figure}
	\label{fig:EMode_0554_L3_M1}
	\centering
	\includegraphics[width=\columnwidth]{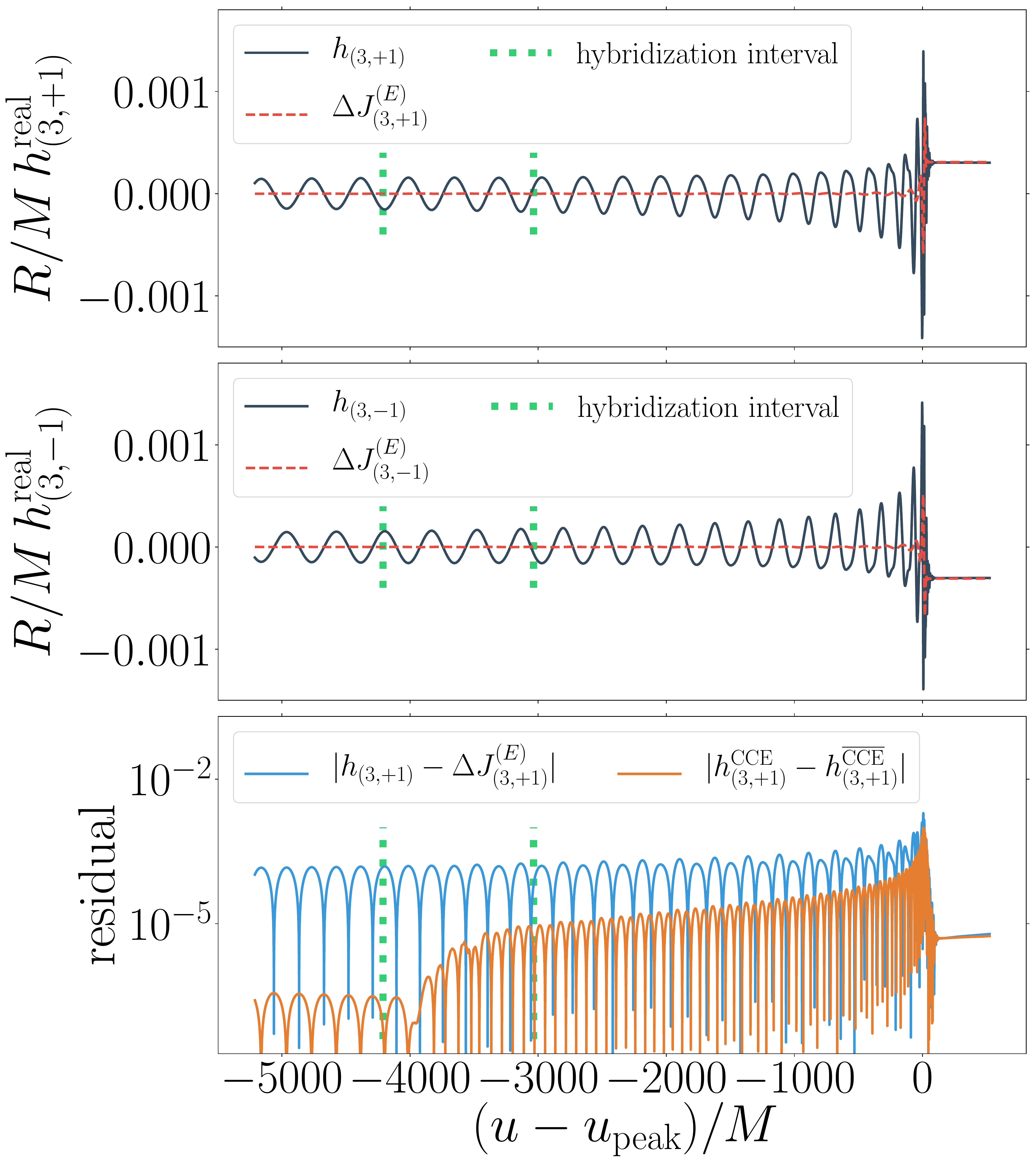}
	\caption{Comparison of the $(3,\pm1)$ modes with the memory for an nonequal mass, nonspinning system (SXS:BBH:0554, see Table~\ref{mergers}).}
\end{figure}

For the other types of binary black hole systems that we examined, the results are very similar to what we have presented thus far except for the following observations. For a nonequal mass, nonspinning system we find that the total accumulated memory is not as large as that occurring in an equal mass system of the same total mass. Furthermore, for a spinning system, we find that the total accumulated memory is constant as a function of spin for antialigned spins, but increases with the total spin for aligned spin systems, which agrees with Ref.~\cite{Pollney_2011}. Also, for a precessing system, we observe mode mixing which causes the electric memory to leak into certain unexpected modes, such as the $(2,1)$ and $(3,0)$ modes. Last, we find that for nonequal mass systems there appears to be memory accumulated in the $(3,\pm1)$ modes, which serves as an example of memory being accumulated in one of the oscillatory modes. We illustrate this effect using SXS:BBH:0554 in Fig.~\ref{fig:EMode_0554_L3_M1}. Although this memory is indeed resolvable relative to numerical error, the value acquired is roughly a third of the total memory that is found in the $(4,0)$ mode and is thus inconsequential when compared to the $(2,0)$ mode's memory, which is nearly two orders of magnitude more than the $(4,0)$ mode's.

Finally, we present Table~\ref{ememories} which contains the memory computed using Eqs.~\eqref{eq:approxemem} and~\eqref{eq:approxbmem} and the memory accumulated in the strain modes, with rough estimates of the corresponding numerical error obtained by comparing the two highest resolution waveforms.
\begin{table*}
	\caption{\label{ememories}%
		Memory values that are obtained by combining Eqs.~\eqref{eq:approxemem} and~\eqref{eq:approxbmem} and those obtained from the overall net change in the extracted strain memory modes. Again, the error that we provide in the final column is simply the residual between the two highest resolution waveforms.}
	\centering
	\begin{ruledtabular}
		\begin{tabular}{ccccccc} 
		SXS:BBH: & $h_{(2,0)}(u_{\text{final}})$ & $\Delta J_{(2,0)}(u_{\text{final}})$ & Error & $h_{(4,0)}(u_{\text{final}})$ & $\Delta J_{(4,0)}(u_{\text{final}})$ &Error \\ [0.5ex] 
		\colrule
		0305 & $9.00\times10^{-2}$ & $8.97\times10^{-2}$ & $1.02\times10^{-5}$ & $1.61\times10^{-3}$ & $1.46\times10^{-3}$ & $4.71\times10^{-5}$\\
		1155 & $9.14\times10^{-2}$ & $9.06\times10^{-2}$ & $5.60\times10^{-6}$ & $1.63\times10^{-3}$ & $1.54\times10^{-3}$ & $2.44\times10^{-6}$\\
		0554 & $7.16\times10^{-2}$ & $7.11\times10^{-2}$ & $6.91\times10^{-6}$ & $8.35\times10^{-4}$ & $7.18\times10^{-4}$ & $1.48\times10^{-5}$\\
		1412 & $9.34\times10^{-2}$ & $9.13\times10^{-2}$ & $2.48\times10^{-4}$ & $1.30\times10^{-3}$ & $1.31\times10^{-3}$ & $9.51\times10^{-6}$\\
		1389 & $6.83\times10^{-2}$ & $6.67\times10^{-2}$ & $5.42\times10^{-3}$ & $7.71\times10^{-4}$ & $7.10\times10^{-4}$ & $2.69\times10^{-4}$\\
		\end{tabular}
	\end{ruledtabular}
\end{table*}

\subsection{Magnetic Memory Modes}
\label{sec:bmemory}

There has been much speculation regarding whether the magnetic part of the displacement memory vanishes, i.e., if $\Delta J^{(B)}=0$ \cite{Flanagan_2017, Winicour_2014, Bieri_2014, ashtekar2019compact}.\footnote{While the magnetic memory $\Delta J^{(B)}$ may indeed vanish, this does not mean that $J^{(B)}(u)$---the magnetic component of the strain--- or even $\Delta J^{(B)}(u)$---the magnetic memory as a function of time--- must be zero, but rather that their overall net change is zero.} As proved by Bieri and Garfinkle~\cite{Bieri_2014}, at linear order, the magnetic part vanishes provided that the news vanishes: $\dot{h}\rightarrow0$ for $u\rightarrow+\infty$. We similarly find that our nonlinear expression as the magnetic memory in terms of the strain's $1/r$ part, i.e., Eq.~\eqref{eq:bmemoryfinal}, also is zero for cases with vanishing news. Unfortunately, confirming that the magnetic component of the memory vanishes in complete generality is not as analytically simple; so, we instead turn to the results of our numerical computations of the magnetic memory.

Unlike the electric memory, which as illustrated earlier is primarily amassed during just the merger phase of a BBH system's coalescence, the magnetic memory as a function of time also acquires meaningful contributions throughout the system's inspiral phase. These contrasting accumulation rates are because of the electric memory's relation to the binary system's energy flux, while the magnetic memory, by contrast, is instead related to the angular momentum flux. As a result, we find that to study accurate magnetic memory effects and observe reasonable agreement between the strain spin memory modes and the spin memory computed from the flux-balance laws, i.e., by calculating $\int\Delta J^{(B)}(u)\,du$, we need to examine numerical simulations with roughly 100 orbits or more. Unfortunately, such simulations are fairly sparse in the SXS Catalog. But as outlined in Table.~\ref{mergers}, there are a few of these $\roughly$100 orbit mergers that we examine now.

By computing the magnetic memory using Eq.~\eqref{eq:approxbmem}, we find that the maximum value of magnetic memory as a function of the angle in the sky for SXS:BBH:1412 is
\begin{align*}
R/M\,\max\big(|\Delta J^{(B)}|(\theta,\phi)\big)=2.31\times10^{-7}\pm2.60\times10^{-2}.
\end{align*}
It is often speculated that a superkick system\footnote{A system with initially antiparallel spins in the orbital plane.} may be the best candidate for producing magnetic memory~\cite{ashtekar2019compact}. For the superkick waveform SXS:BBH:0963,\footnote{The relevant parameters of this system are 
\begin{align}
M_{1}/M_{2}&=1.0,\quad N_{\text{orbits}}=19,\nonumber\\
\chi_{1}^{\text{initial}}&=(+0.18, -0.78, -1.2\times10^{-3}),\nonumber\\
\chi_{2}^{\text{initial}}&=(-0.16,+0.78,+1.2\times10^{-3}).
\end{align}}
we find
\begin{align*}
R/M\,\max\big(|\Delta J^{(B)}|(\theta,\phi)\big)=9.37\times10^{-5}\pm1.75\times10^{-2}.
\end{align*}
Therefore, the magnetic memory is consistent with zero \cite{Flanagan_2017, Winicour_2014, Bieri_2014, ashtekar2019compact}.

Because the magnetic memory effect for each system we have looked at is much smaller than the corresponding numerical error, we believe that we are most likely overestimating the magnetic memory's numerical uncertainty. While the magnetic component of the memory appears to be zero, we expect the spin memory, i.e., the retarded time integral of the magnetic memory, to take on some nonzero final value in a manner similar to that of the electric memory. Because of this, we only provide one example of a magnetic memory mode and reserve a more exhaustive presentation for the nonzero spin memory, which we examine in Sec.~\ref{sec:spinmemory}.

\begin{figure}
	\includegraphics[width=\columnwidth]{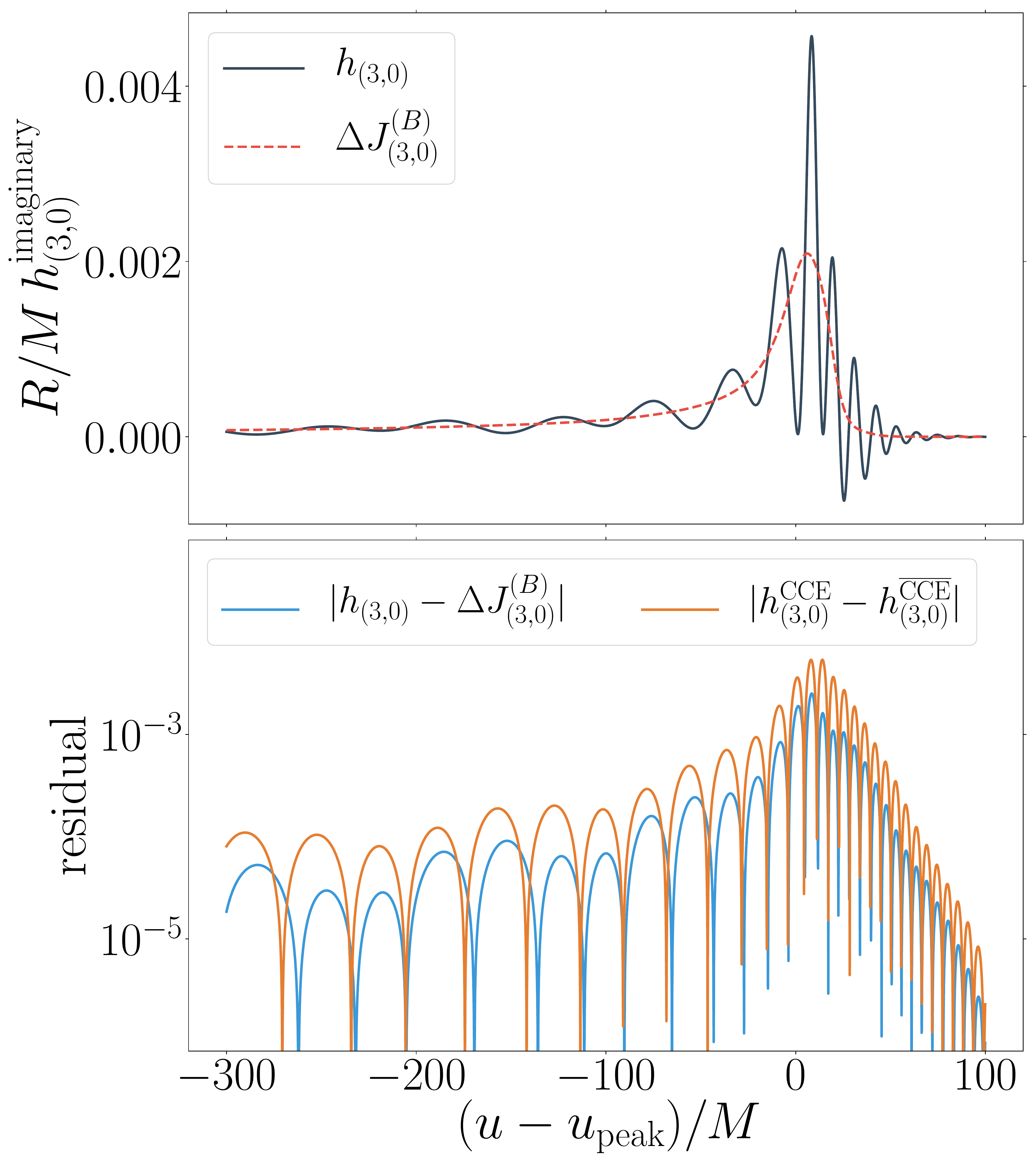}
	\caption{Comparison of imaginary part of the $(3,0)$ mode with the magnetic memory for a $N_{\text{orbit}}\approx150$ system (SXS:BBH:1412, see Table~\ref{mergers}).}
	\label{fig:BMode_1412_M0}
\end{figure}

From earlier comparisons with PN approximations~\cite{Nichols_2017}, we expect the primary magnetic memory contributions to be from the imaginary part of the $\ell=\text{odd}$, $m=0$ modes, with the most pronounced mode being the $(3,0)$ mode. In Fig.~\ref{fig:BMode_1412_M0} we compare the most prominent strain magnetic memory mode to the computed magnetic memory.
\noindent
Notice, not unlike the electric memory, the magnetic memory tends to act as the average of the more oscillatory strain. While the $(3,0)$ mode may seem to be poorly resolved near the system's merger phase, this is merely a consequence of examining SXS's $\roughly$100 orbit runs, whose available numerical resolutions tend to be poorer than the other runs in the SXS Catalog. One can easily observe this by examining the $(3,0)$ mode shown in Fig.~\ref{fig:ExtComparison}, which shows this mode for SXS:BBH:0305: a run with a much more accurate and precise Cauchy evolution.

\subsection{Spin Memory Modes}
\label{sec:spinmemory}
\begin{figure}
	\includegraphics[width=\columnwidth]{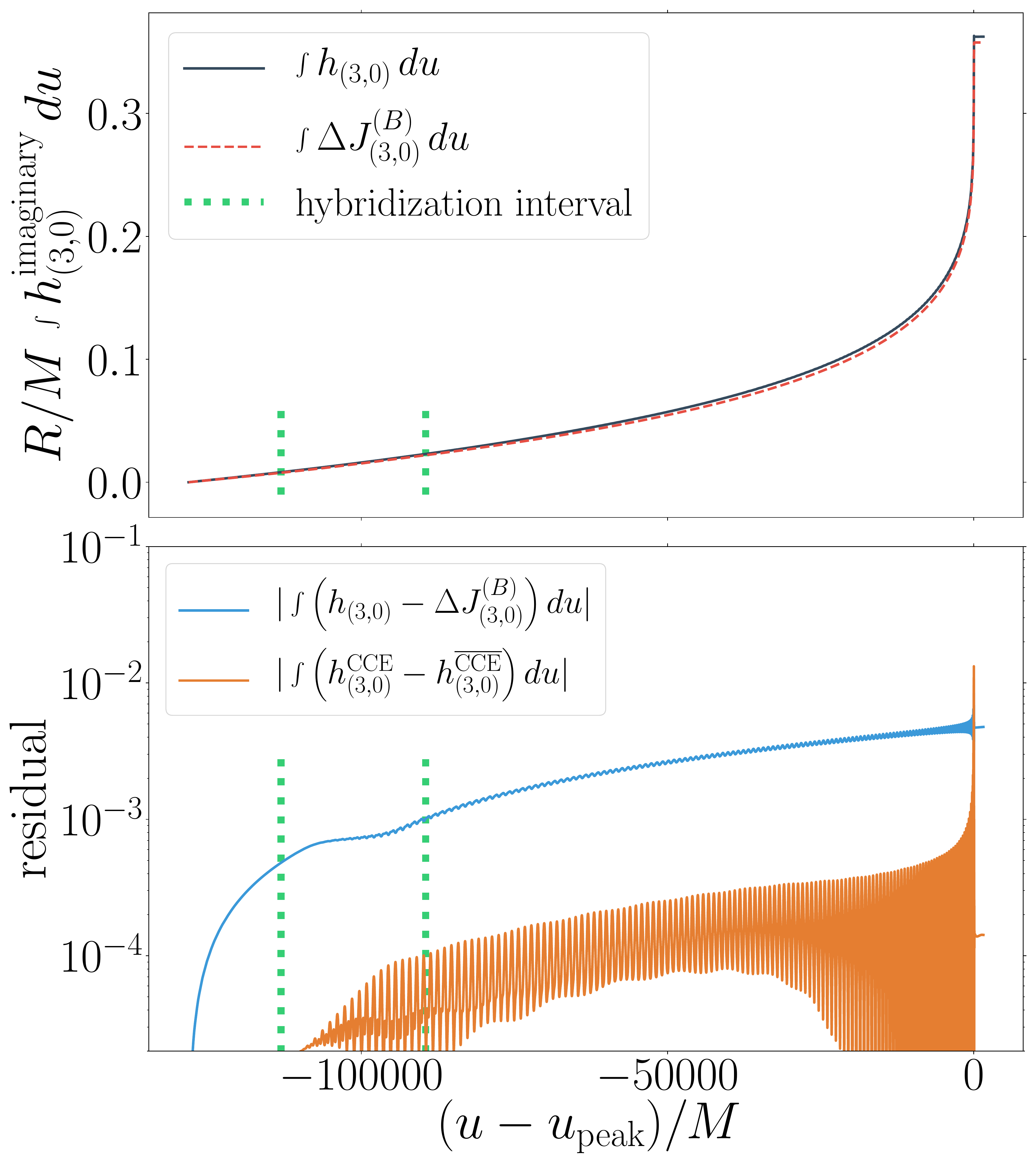}
	\caption{Comparison of the retarded time integral of the imaginary part of the $(3,0)$ mode with the spin memory for a $N_{\text{orbit}}\approx150$ system (SXS:BBH:1412, see Table~\ref{mergers}).}
	\label{fig:SMode_1412_M0}
\end{figure}

We now evaluate the spin memory $\int\Delta J^{(B)}(u)\,du$, which we compute by taking the time integral of Eq.~\eqref{eq:approxbmem}. Because the spin memory, as with the magnetic memory, corresponds to the angular momentum flux, we expect the spin memory to closely resemble the electric memory, but with a considerably larger build-up during inspiral. As we show in Fig.~\ref{fig:SMode_1412_M0}, this is the case as nearly the same amount of spin memory is accumulated throughout the system's inspiral phase as there is in the merger phase.
\noindent
Further, like the electric memory and its $(4,0)$ mode, we find that we can also resolve the next most prominent spin memory mode---namely, the $(5,0)$ mode---to within numerical error, but not the other $m=0$ modes.

Last, we present Table~\ref{spinmemories}, which is of the same form as Table~\ref{ememories}, but contains the values of the spin memory computed by integrating Eq.~\eqref{eq:approxbmem} and the spin memory found in the retarded time integral of the strain modes.

\begin{table*}
	\caption{\label{spinmemories}%
		Spin memory values obtained by computing the retarded time integral of Eq.~\eqref{eq:approxbmem} and those obtained from the overall net change in the retarded time integral of the extracted strain spin memory modes. Again, the error that we provide in the final column is simply the residual between the two highest resolution waveforms.}
	\centering
	\begin{ruledtabular}
		\begin{tabular}{ccccccc} 
		SXS:BBH: & $\int^{u_{\text{final}}}h_{(3,0)}(u)\,du$ & $\int^{u_{\text{final}}}\Delta J_{(3,0)}^{(B)}(u)\,du$ & Error & $\int^{u_{\text{final}}}h_{(5,0)}(u)\,du$ & $\int^{u_{\text{final}}}\Delta J_{(5,0)}^{(B)}(u)\,du$ & Error\\ [0.5ex] 
		\colrule
		0305 & $4.05\times10^{-1}$ & $3.61\times10^{-1}$ & $7.24\times10^{-5}$ & $8.56\times10^{-4}$ & $9.53\times10^{-4}$ & $1.22\times10^{-5}$\\
		1155 & $4.32\times10^{-1}$ & $3.55\times10^{-1}$ & $1.53\times10^{-4}$ & $1.09\times10^{-3}$ & $1.03\times10^{-3}$ & $5.85\times10^{-6}$\\
		0554 & $3.28\times10^{-1}$ & $2.85\times10^{-1}$ & $1.21\times10^{-5}$ & $1.80\times10^{-4}$ & $2.15\times10^{-4}$ & $1.70\times10^{-5}$\\
		1412 & $3.62\times10^{-1}$ & $3.58\times10^{-1}$ & $1.42\times10^{-4}$ & $7.06\times10^{-4}$ & $7.46\times10^{-4}$ & $1.39\times10^{-6}$\\
		1389 & $2.79\times10^{-1}$ & $2.88\times10^{-1}$ & $4.13\times10^{-2}$ & $3.12\times10^{-4}$ & $3.64\times10^{-4}$ & $6.92\times10^{-5}$\\
		\end{tabular}
	\end{ruledtabular}
\end{table*}

\subsection{Fitting Ringdown to QNMs}
\label{sec:qnms}
We now investigate the oscillatory ringdown part of the $(2,0)$ and $(3,0)$ modes, which otherwise correspond to the electric and magnetic memory. We wish to explain the ringdown part of these modes with perturbation theory, i.e., by fitting them to the expected quasinormal modes. As was recently explored by Giesler \emph{et al.}~\cite{Giesler_2019}, once a BBH system has merged into a single black hole, the resulting black hole ringdown is well described by a \emph{linear} superposition of quasinormal modes even from as early as the peak of the waveform, provided that the overtones are included. These quasinormal modes can be used to find the mass and spin angular momentum of the final black hole~\cite{PhysRevD.75.124018, PhysRevD.97.044048,Giesler_2019}. Thus far, though, only the $(2,2)$ mode has been thoroughly examined. Consequently, while we do not attempt to estimate the final black hole's characteristics using our fits to the $(2,0)$ and $(3,0)$ modes, we nonetheless present the accuracy of our fits, saving the parameter estimation and analysis for a future work.

Like previous work on quasinormal modes~\cite{Vishveshwara_1970, Press_1971, Teukolsky_1973, Chandrasekhar_1975}, we model the radiation occurring during ringdown as a sum of damped sinusoids with complex frequencies $\omega_{lmn}=\omega_{lmn}(M_{f},\chi_{f})$ which can be computed by using perturbation theory~\cite{Stein_2019}. But, because the strain now exhibits memory effects that are not captured by the usual quasinormal mode expression, we instead perform a superposition of the memory and the quasinormal modes:
\begin{align}
\label{eq:qnmmodel}
h_{lm}^{N}=\Delta J(u)+\sum\limits_{n=0}^{N}C_{lmn}e^{-i\omega_{lmn}(u-u_0)}\quad u\geq u_0,
\end{align}
\begin{figure*}
	\label{fig:QNM}
	\centering
	\includegraphics[width=\columnwidth]{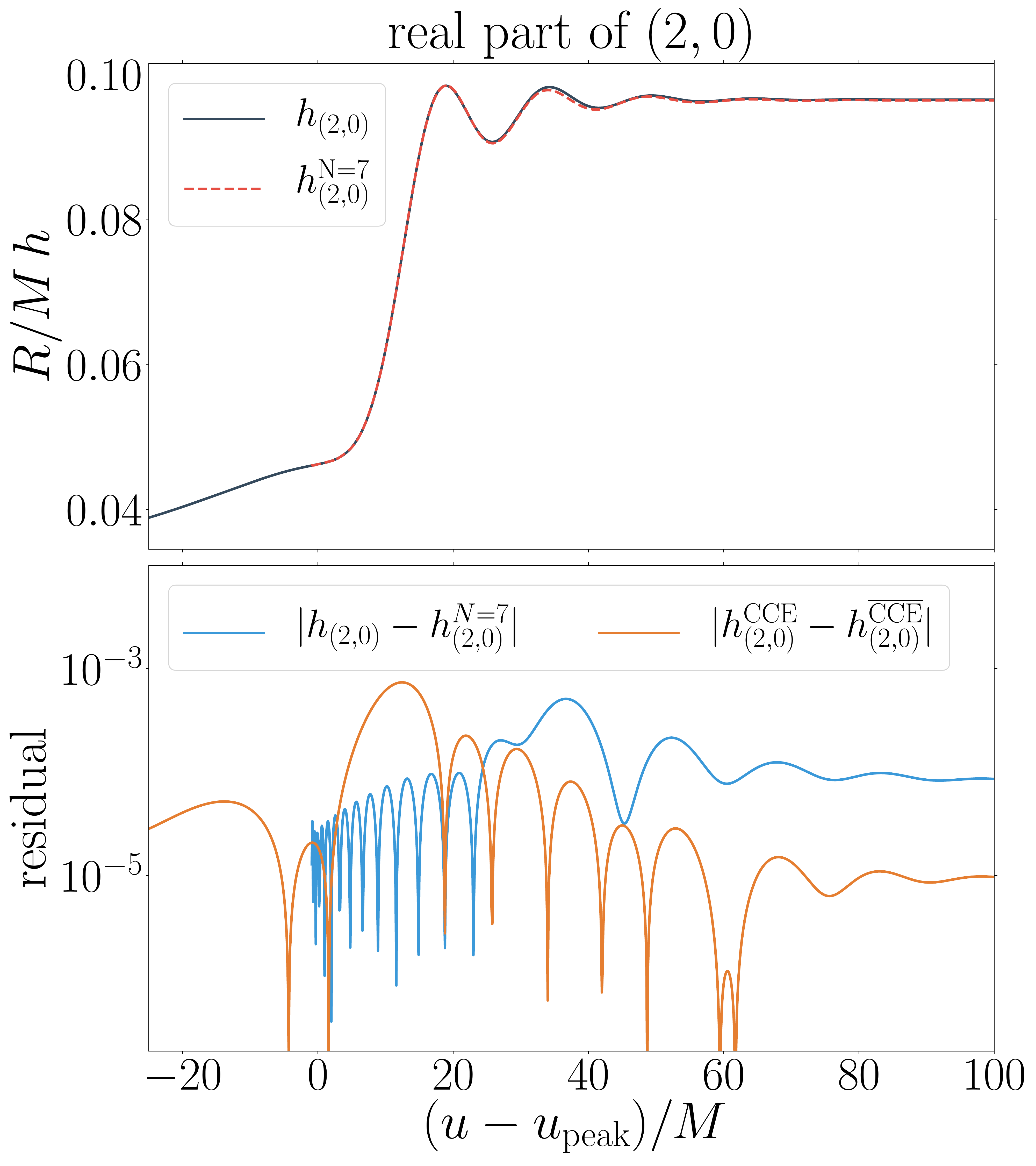}
	\hspace{0.2cm}
	\includegraphics[width=\columnwidth]{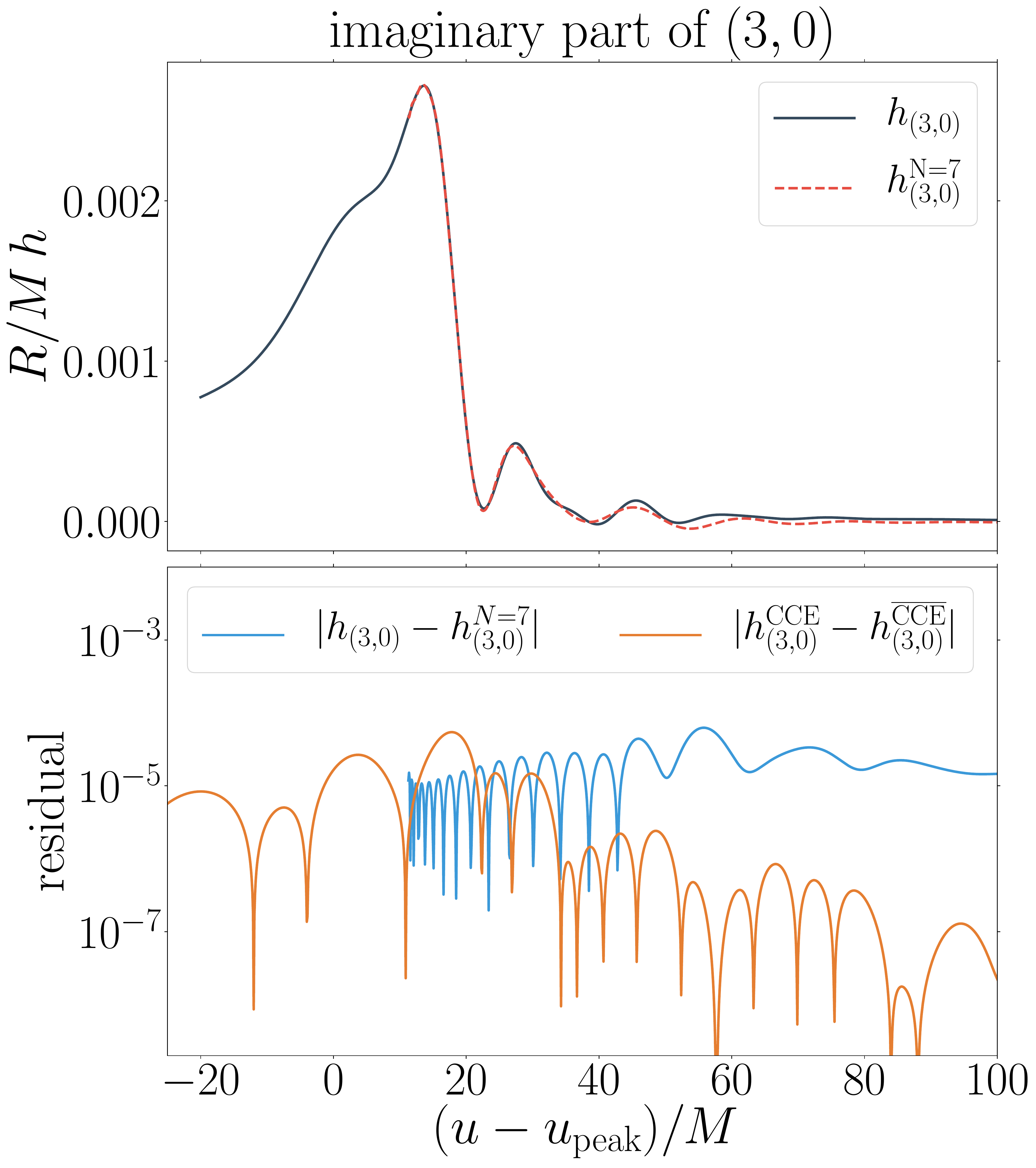}
	\caption{Comparison between the numerical relativity waveform and the $N=7$ ``QNM + memory'' model for the primary electric and magnetic memory modes of SXS:BBH:0305. We start the QNM model at $u_0\approx0\,M$ for the $(2,0)$ mode and at $u_{0}\approx10\,M$ for the $(3,0)$ mode. The top row shows the strain and its corresponding fit, while the bottom row shows the residual. We also show an estimate of the error in the CCE waveform, $|h_{(\ell,m)}^{\text{CCE}}-h_{(\ell,m)}^{\overline{\text{CCE}}}|$, where $h_{(\ell,m)}^{\text{CCE}}$ refers to the highest resolution waveform of SXS:BBH:0305 and $h_{(\ell,m)}^{\overline{\text{CCE}}}$ refers to the next highest resolution.}
\end{figure*}\noindent
where $N$ is the number of overtones used in our fitting and $u_{0}$ is a specifiable ``start time'' for the model, with any times that occur before $u_{0}$ not being included in the fits. Recall that in this paper we approximate $\Delta J(u)$ with only the null memory, ignoring the ordinary memory; this may introduce some error in our fits to Eq.~\eqref{eq:qnmmodel}. However, since the ordinary part's contribution is fairly minor---roughly 0.3\% that of the null part's contribution---our fits to the QNMs should be reasonably accurate. Further, because the QNM expressions tend to zero as \(u\rightarrow+\infty\), rather than making the strain and the memory be equal at their initial values, we instead make them coincide at the time \(u_{\text{final}}\). With our adjusted waveforms, we then fit Eq.~\eqref{eq:qnmmodel} to the $(2,0)$ and $(3,0)$ modes.

We construct fits for the simulation SXS:BBH:0305. We find the mismatch
\begin{align}
\mathcal{M}=1-\frac{\langle h_{\ell m},h_{\ell m}^{N}\rangle}{\sqrt{\langle h_{\ell m},h_{\ell m}\rangle\langle h_{\ell m}^{N},h_{\ell m}^{N}\rangle}}.
\end{align}
 between our fits and the memory modes are minimized for $u_{0}\approx0\,M$ for the $(2,0)$ mode, while an initial time of $u_{0}\approx10\,M$ is needed to minimize the mismatch for the $(3,0)$ mode. We believe that the $(3,0)$ mode likely needs a larger value of $u_{0}$ because the error in that mode is larger than that of the $(2,0)$ mode, so the magnetic memory is not as accurate and thus the QNM model needs to start further on in the ringdown phase to minimize the effect of this inaccuracy. In Fig.~\ref{fig:QNM} we present the fit results for the simulation SXS:BBH:0305 at the optimal fit times $u_{0}$ as found by minimizing the corresponding mismatch between the strain and the fit. The final mismatches for these modes are then
 \begin{align*}
 \mathcal{M}\big(\text{Re}(h_{2,0})\big)&=4.01\times10^{-7},\\
 \mathcal{M}\big(\text{Im}(h_{3,0})\big)&=6.57\times10^{-4}.
 \end{align*}

\subsection{Signal-to-Noise Ratios}
\label{sec:snrs}

We now investigate the measurability of the memory by calculating the signal-to-noise ratios for the displacement and spin memory effects in a few of the current and planned GW detectors. We compute the SNR $\rho$ using
\begin{align}
\rho=\sqrt{4\int_{f_{\text{min}}}^{f_{\text{max}}}\frac{|\tilde{h}(f)|^2}{S_{n}(f)}\,df},
\end{align}
where $\tilde{h}(f)$ is the Fourier transform in frequency of the detector response $h(u)$ (see Eq.~\eqref{eq:detectorresponse}), $S_{n}(f)$ is the noise power-spectral density (PSD), and $f_{\text{min}}$ and $f_{\text{max}}$ are frequency limits that are regulated by the chosen PSD. We construct $h(u)$ as
\begin{align}
\label{eq:detectorresponse}
h(t)&=F_{+}(\theta,\phi,\psi)h_{+}(u,\iota,\phi_{0})\nonumber\\
&\phantom{=.}+F_{\times}(\theta,\phi,\psi)h_{\times}(u,\iota,\phi_{0}),
\end{align}
where $F_{+}$ and $F_{\times}$ are the antenna response patterns,
\begin{subequations}
\begin{align}
F_{+}&=\frac{1}{2}\big(1+\cos^2\theta\big)\cos(2\phi)\cos(2\psi)\nonumber\\
&\phantom{=.}-\cos\theta\sin(2\phi)\sin(2\psi),\\
F_{\times}&=\frac{1}{2}\big(1+\cos^2\theta\big)\cos(2\phi)\sin(2\psi)\nonumber\\
&\phantom{=.}+\cos\theta\sin(2\phi)\cos(2\psi),
\end{align}
\end{subequations}
with $\theta$ and $\phi$ being the spherical coordinates relative to the observatory's axes and $\psi$ the angle between the two usual polarization components $h_{+}$ and $h_{\times}$ and the observatory's two axes. The angles $\iota$ and $\phi_{0}$ are the spherical coordinates relative to the BBH's source frame. While these angles could take on a variety of values, to simplify our computations we choose the values that maximize the SNR for the respective memory observables.

\begin{figure}
	\includegraphics[width=\columnwidth]{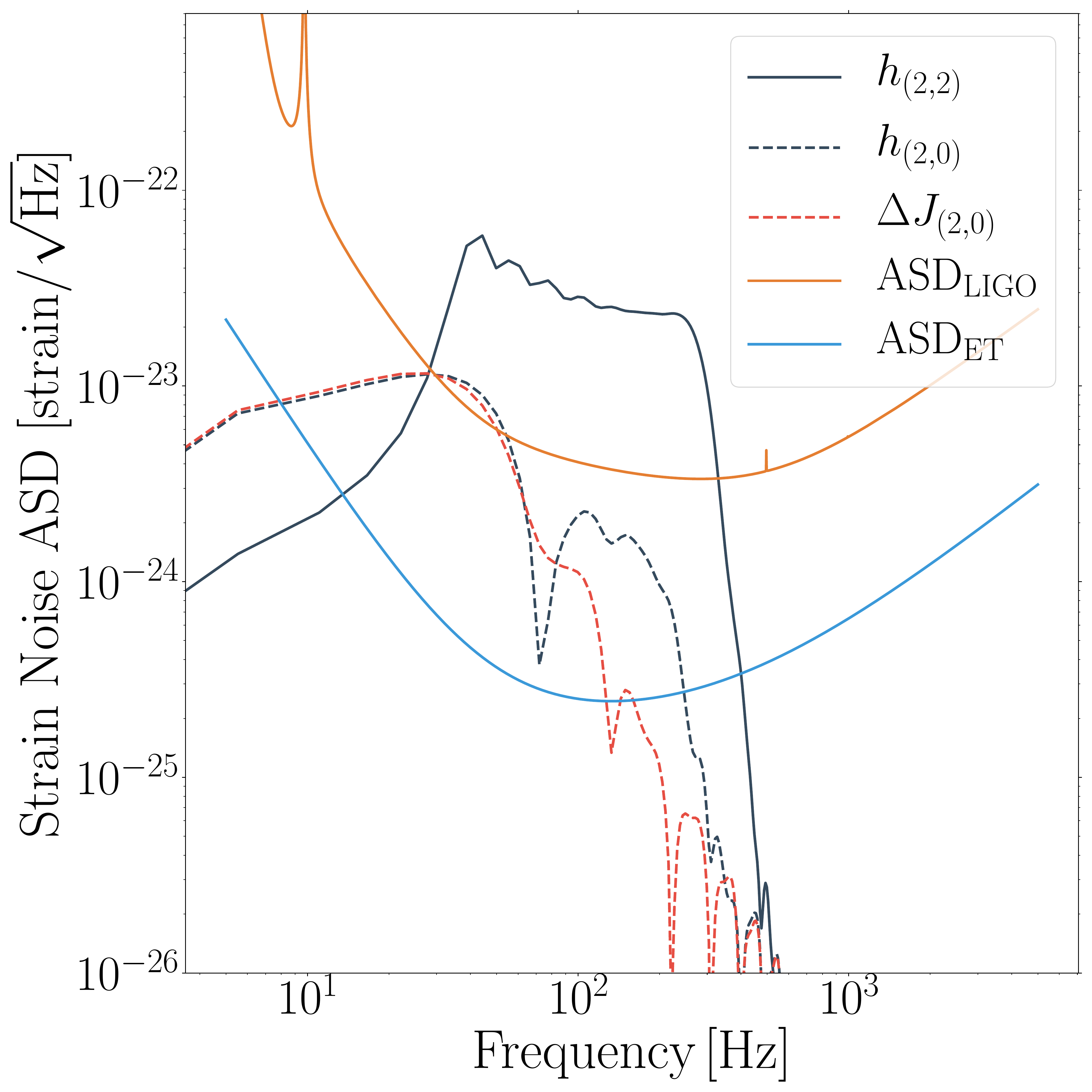}
	\caption{Both of LIGO's and the Einstein Telescope's (ET) amplitude spectral densities (ASD) compared to the strain $(2,2)$ (black/solid) and $(2,0)$ (black/dashed) modes as well as the memory's $(2,0)$ mode (red/dashed).}
	\label{fig:memfrequency}
\end{figure}

We examine SNRs for LIGO, the Einstein Telescope,\footnote{Specifically, the single-interferometer configuration (ET-B).} and LISA using the simulation SXS:BBH:0305, which for the values $M=65\,M_{\odot}$ and $R=410\,\text{Mpc}$ resembles the first event that was observed by LIGO: GW150914~\cite{Abbott_2016}. When computing the LISA SNRs, though, we instead use the mass \(M=10^{5}\,M_{\odot}\) to mimic the mass of supermassive black hole binaries, which places the memory signal near the bucket of the LISA noise curve. For LIGO SNRs, we use the updated \href{https://dcc.ligo.org/LIGO-T1800044/public}{Advanced LIGO sensitivity design curve}, while for the ET and LISA SNRs we use the the sensitivity curve approximations that are shown in Eq.~(19) of \cite{Regimbau_2012} and Eq.~(1) of ~\cite{Robson_2019}. For our SNRs, we only examine the primary electric and magnetic modes because the other modes' contributions are negligible. Furthermore, we find that it is important to only consider the null memory when computing SNRs, rather than the strain, because the QNM frequencies in the strain can contaminate and thus obscure the memory SNRs, as is illustrated in Fig.~\ref{fig:memfrequency}. In other words, the $h_{(2,0)}$ mode contains higher frequencies due to QNM oscillations than the $\Delta J_{(2,0)}$ mode, which just describes the growth of the memory, and will thus yield a larger memory SNR than the true memory SNR.

In Table~\ref{SNRs}, we present the results that we find for these orientation-optimized SNRs. Alongside the SNRs for the $(2,0)$ and $(3,0)$ modes, we also provide the SNR for the $(2,2)$ mode computed using the same orientation that was chosen for the specific memory mode.

\begin{table}[H]
\caption{\label{SNRs}%
LIGO, ET, and LISA SNRs for the most prominent electric and magnetic memory
modes from SXS:BBH:0305. The LIGO and ET SNRs are for a total mass of
$M=65\,M_{\odot}$, while for LISA we use $M=10^{5}\,M_{\odot}$.}
\begin{ruledtabular}
    \begin{tabular}{ccccc}
        Detector & $\Delta h_{(2,0)}$ & $h_{(2,2)}$ & $\Delta h_{(3,0)}$
           & $h_{(2,2)}$  \\ [0.5ex]
        \colrule
        LIGO & $2.12\times10^{+0}$ & $2.03\times10^{+1}$ &
               $6.36\times10^{-2}$ & $5.06\times10^{+1}$ \\
        ET-B & $3.48\times10^{+1}$ & $3.14\times10^{+2}$
             & $1.05\times10^{+0}$ & $7.83\times10^{+2}$ \\
        LISA & $1.44\times10^{+2}$ & $2.98\times10^{+2}$
             & $3.16\times10^{+0}$ & $7.49\times10^{+2}$ \\
    \end{tabular}
\end{ruledtabular}
\end{table}

The SNRs that we find are larger, even if only slightly, than previous calculations that have taken advantage of either PN or postprocessing methods. Because of this, we conclude that the memory effect will most likely only be measured in future observatories or by stacking signals recorded by LIGO for $\roughly$100 events, which should take about five years~\cite{Lasky_2016, Boersma_2020}.

\subsection{Cauchy-Characteristic Extraction}
\label{sec:cce}
Finally, we discuss some of the important features of $\texttt{SpECTRE}$'s Cauchy-characteristic extraction that need to be dealt with to successfully extract memory effects. As has been remarked by Favata and others~\cite{Favata_2009_PN, Favata_2010, Nichols_2017}, as well as in Fig.~\ref{fig:ExtComparison}, numerical relativity simulations that employ Reggi-Wheeler-Zerilli perturbative extraction or Newman-Penrose (NP) extraction have so far been unable to resolve the $m=0$ modes, which contain the majority of the memory effect induced by a BBH merger.
Currently, the reason for this issue is not known. Fortunately, though, Cauchy-characteristic extraction~\cite{barkett2019spectral} can succeed.

Unlike the RWZ and NP extraction schemes, CCE takes the finite-radius world tube information created by a Cauchy evolution as the inner boundary data for an evolution of Einstein's field equations on hypersurfaces constructed by constant retarded time. Consequently, gravitational waves can then be computed directly from Einstein's equations at future null infinity. Further, since $\texttt{SpECTRE}$'s CCE extracts the strain independently of the news, unlike previous works that have used CCE~\cite{Pollney_2011}, there is no need to integrate the news with respect to retarded time, which introduces ambiguities because of an unknown integration constant.

Despite the improved precision of the CCE waveforms, there is a degree of freedom in the procedure that needs to be dealt with. The characteristic evolution within CCE allows one field, the strain $h$, to be (almost) freely chosen on the initial null hypersurface, and the choice of that field then influences the waveform at future null infinity. Consequences of this choice manifest as transient effects that appear at early retarded times. We can eliminate these effects by choosing a late enough transition time when hybridizing the CCE strain with the PN waveform. The transient effects caused by the choice of $h$ on the initial null hypersurface were previously explored in \cite{Bishop_2016}. For this paper, we choose initial data to match the value and first radial derivative of $h$ from the Cauchy data on the world tube, using the simple ansatz
\begin{equation}
\label{eq:inverse_cubic}
h(u=0,r,\theta^A) = \frac{A(\theta^A)}{r} + \frac{B(\theta^A)}{r^3}.
\end{equation}
The two coefficients $A(\theta^A)$ and $B(\theta^A)$ are fixed by the Cauchy data on the world tube. The form of Eq.~\eqref{eq:inverse_cubic} is chosen to maintain regularity of the characteristic system, which requires a careful choice of gauge and initial data in which the $\propto 1/r^2$ part vanishes at future null infinity.

\begin{figure}
	\label{fig:ExtractionRComparison}
	\centering
	\includegraphics[width=\columnwidth]{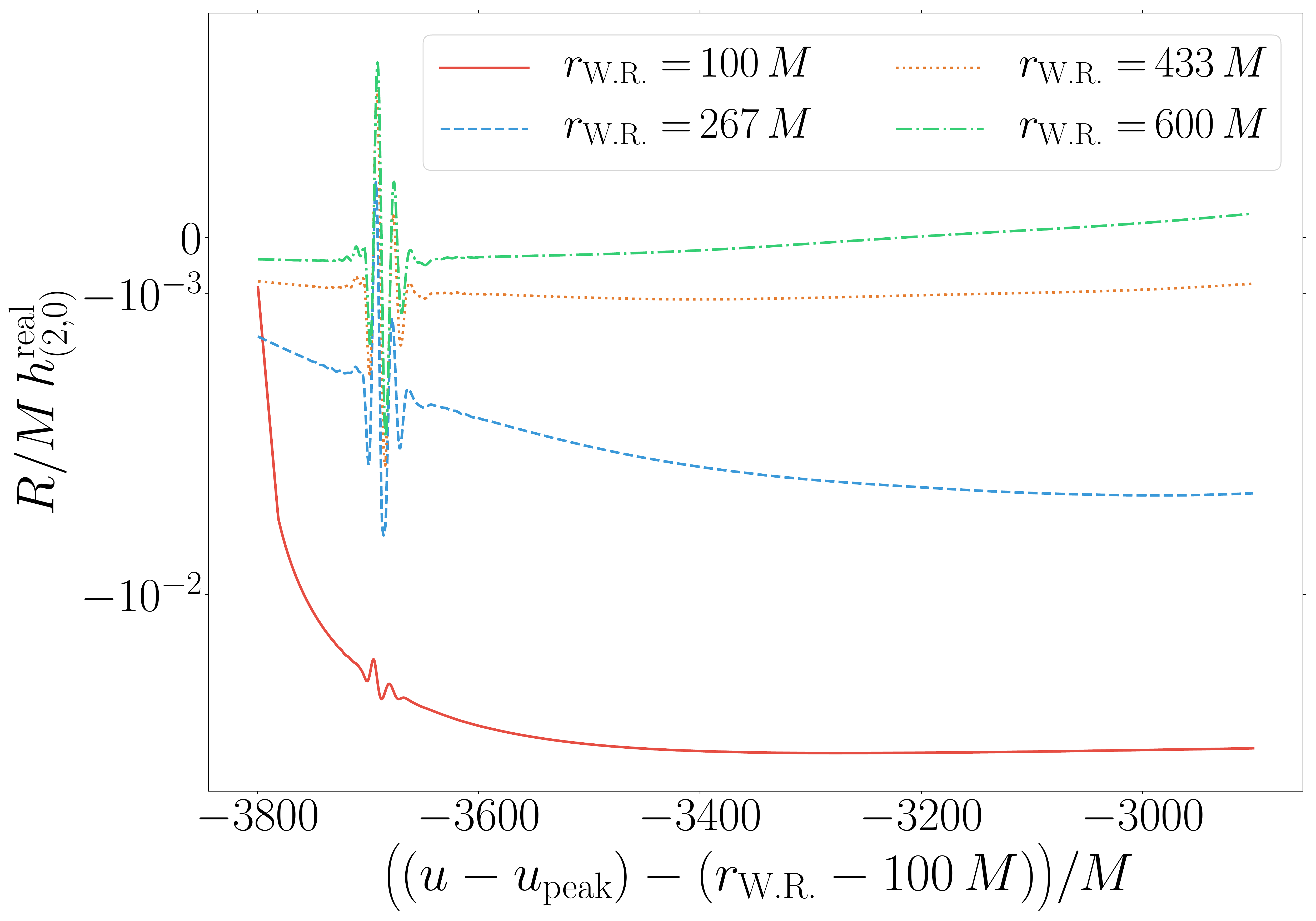}
	\caption{The strain $(2,0)$ mode computed using CCE world tubes of various radii $r_{\text{W.R.}}$, for the simulation SXS:BBH:0305. We have added a time translation so that waveforms for every $r_{\text{W.R.}}$ begin at the same time. We show only the beginning of the waveform, and the values here are much smaller than the overall scale of the $(2,0)$ mode as seen, e.g., in Fig.~\ref{fig:PNComparison_L2_M0}.}
\end{figure}

As we illustrate in Fig.~\ref{fig:ExtractionRComparison}, the initial behavior of the $(2,0)$ mode of the strain is dependent upon the choice of the world tube radius that one makes: a smaller radius results in the strain becoming more negative once the junk passes. Similar to the junk radiation seen around $-3700\,M$ in Fig.~\ref{fig:ExtractionRComparison}, the initial transient radiation in CCE is a result of numerical relativity not possessing a complete past history of the binary system's evolution. Fortunately, we find that we can remedy this junk effect by constructing a numerical relativity and PN hybrid, which starts at a time that corresponds to $4$ times the worldtube radius, e.g., $u\approx 400\,M$ for $r_{\text{W.R.}}=100\,M$, and extends throughout $\roughly20\%$ of the numerical waveform. For the results we presented earlier, we chose to hybridize the second smallest world tube radius waveform, seeing as this waveform produced the best agreement between the strain and the BMS flux-balance strain.

\section{Conclusion}
\label{sec:conclusion}
When a binary black hole merger emits radiation that propagates through spacetime toward asymptotic infinity, persistent physical changes known as memory effects occur. These changes are induced as a consequence of BMS flux-balance laws that extend the Poincar\'e balance laws. Because these BMS flux-balance laws physically relate to supertranslations, -rotations, or -boosts, these changes are called displacement, spin, or center-of-mass memories. Measuring these memory effects will be an important test of Einstein's theory of general relativity. However, computing the memory produced in a binary black hole merger requires numerical relativity. Before this work, studying memory using numerical relativity has been challenging because many of the memory contributions to the metric could not be properly resolved.

Using a collection of energy and angular momentum flux equations, we computed the  memory that is induced in vacuum spacetimes as a function of the radiated strain, thereby allowing for the complete calculation of both the electric and magnetic components of the memory effect. We then verified that the strain and the two Weyl scalars from $\texttt{SpECTRE}$'s Cauchy-characteristic extraction obey the two BMS flux-balance equations that we used to calculate the memory. While performing this check, we saw that the primary contribution to the memory comes from the null contribution, because the simulations in the SXS Catalog tend to experience no supertranslations or super-Lorentz transformations. We derived an expression for the null memory contribution that depends on every one of the strain modes and on some of the Weyl scalars. We compared this expression with the well-understood $m=0$ memory modes of the strain, for many simulations of BBH mergers spanning a variety of input parameters. Overall, this new expression for the memory effect agrees with the strain very well, and our comparison works even for the conjectured oscillatory $m\not=0$ memory modes. Furthermore, we found that the magnetic component of the memory, which is believed to be zero, indeed vanishes to the precision of the corresponding Cauchy simulation.

In addition, we found that we can rather accurately model the various memory modes as the combination of a memory signal during the inspiral and merger phases and a quasinormal mode signal during the ringdown phase. Our best fits to the two primary electric and magnetic memory modes offer the possibility that memory modes could participate in constraining the remnant black hole's mass and spin. However, the extent to which the inclusion of the memory modes can improve parameter estimation remains a subject for a future investigation.

Last, we found that the memory SNRs for LIGO, the Einstein Telescope, and LISA are slightly better than previous expectations. Consequently, memory should be observable with future detectors or once a big enough catalog of merger events is obtained by LIGO.

During the past few years, the memory effect was shown to be equivalent to Weinberg's soft theorem through a Fourier transform in time~\cite{Strominger_2014, Pasterski_2016}, thus forming a curious connection between memory, asymptotic symmetries, and soft theorems. Because of this, memory can perhaps serve as an important physical realization of these abstractly formulated results, and thus may one day help realize the holographic structure of quantum gravity in arbitrary four-dimensional spacetimes.

\section*{ACKNOWLEDGMENTS}
We thank Matt Giesler for many useful conversations. Computations were performed on the Wheeler cluster at the California Institute of Technology (Caltech), which is supported by the Sherman Fairchild Foundation and by Caltech. This work was supported in part by the Sherman
Fairchild Foundation and by NSF Grants No. PHY-1708212, No. PHY-1708213, and No. OAC-1931266 at
Caltech and NSF Grants No. PHY-1912081 and No. OAC-1931280 at Cornell.

\appendix

\section{BONDI MASS ASPECT}
\label{a:bondimass}
As was shown in Sec.~\ref{sec:ememorycomputation}, the ordinary contribution to the electric component of the memory is a function of the Bondi mass aspect \(m\). Consequently, to compute the electric memory from numerical relativity waveforms, one needs to know the Bondi mass aspect in terms of the strain and the Weyl scalar \(\Psi_{2}\). Using the results that were obtained by Moxon \emph{et al.}~\cite{moxon2020improved}, by rearranging their Eq.~(94e) and converting their notation to ours, we find
\begin{align}
\label{eq:bondimass}
m=-\text{Re}\left[\Psi_{2}+\frac{1}{4}\dot{h}\bar{h}\right].
\end{align}
The notation changes that are needed to convert from Moxon's work to ours are \(\mathring{W}^{(2)}\rightarrow-2m\) and \(\mathring{J}^{(1)}\rightarrow\bar{h}\), since Moxon takes \(\mathring{J}^{(1)}\) to have spin-weight $+2$ rather than spin-weight $-2$, which is our convention.

\section{BONDI ANGULAR MOMENTUM ASPECT}
\label{a:bondiangmom}
As was shown in Sec.~\ref{sec:ememorycomputation}, the ordinary contribution to the magnetic component of the memory is a function of the angular momentum aspect \(\widehat{N}_{A}\). Thus, to compute the magnetic memory from numerical relativity waveforms, one needs to know the angular momentum aspect in terms of the strain and the Weyl scalar \(\Psi_{1}\). We start by contracting the $\mathcal{O}(r^{-3})$ part of Eq.~\eqref{eq:metricUterm} with $q_{A}$, from which we obtain
\begin{align}
\label{eq:contractedU}
\mathcal{U}^{(3)}&=-\frac{2}{3}N+\frac{1}{16}\eth(C_{AB}C^{AB})+\frac{1}{2}q_{A}C^{AB}D^{C}C_{BC}.
\end{align}
Using $C_{AB}C^{AB}=2h\bar{h}$ and $C^{AB}D^{C}C_{BC}=\text{Re}\big[q^{A}h\bar{\eth}\bar{h}\big]$ [from Eq.~\eqref{eq:C_{AB}D^{C}N^{BC}}], we can then rewrite Eq.~\eqref{eq:contractedU} as
\begin{align}
\mathcal{U}^{(3)}=-\frac{2}{3}N+\frac{1}{8}\eth(h\bar{h})+\frac{1}{2}\bar{h}\eth h.
\end{align}
According to Bishop \emph{et al.}~\cite{Bishop_1997} Eqs.~(8) and~(A2)
\begin{align}
\label{eq:bishopeq}
\partial_{r}\mathcal{U}=\frac{e^{2\beta}}{r^2}(KQ-\bar{h}\bar{Q})
\end{align}
for
\begin{align}
K\equiv\frac{1}{2}q^{A}\bar{q}^{B}\gamma_{AB}\quad\text{and}\quad Q\equiv q_{A}r^2e^{-2\beta}\gamma^{AB}\partial_{r}\mathcal{U}_{B}.
\end{align}
Thus, by examining the \(\mathcal{O}(r^{-3})\) part of Eq.~\eqref{eq:bishopeq}, we find
\begin{align}
-3\mathcal{U}^{(3)}&=K^{(0)}Q^{(2)}-\bar{h}\bar{Q}^{(1)}+2\beta_{0}K^{(0)}Q^{(1)}\nonumber\\
&=Q^{(2)}-\bar{h}\bar{Q}^{(1)},
\end{align}
seeing as \(\beta_{0}=0\) by Flanagan and Nichols's Eq.~(2.9b)~\cite{Flanagan_2017}. Also, by explicit calculation and Flanagan and Nichols's Eq.~(2.9a), since \(q_{A}D_{B}C^{AB}=\bar{\eth}\bar{h}\), we can write $\bar{Q}^{(1)}$ as
\begin{align}
\label{eq:q1}
\bar{Q}^{(1)}=-2\bar{U}=\eth h.
\end{align}
Furthermore, by Moxon \emph{et al.'s}~\cite{moxon2020improved} Eq.~(94c)
\begin{align}
\Psi_{1}=-\frac{3}{2}\eth\beta_{1}+\frac{1}{8}\bar{h}\bar{Q}^{(1)}+\frac{1}{4}Q^{(2)}.
\end{align}
But, since Flanagan and Nichols's~\cite{Flanagan_2017} Eq.~(2.9c) implies
\begin{align}
\beta_{1}=-\frac{1}{32}C_{AB}C^{AB}=-\frac{1}{16}h\bar{h},
\end{align}
we then have
\begin{align}
\label{eq:q2}
Q^{(2)}=4\Psi_{1}-\frac{3}{8}\eth(h\bar{h})-\frac{1}{2}\bar{h}\bar{Q}^{(1)}.
\end{align}
Combining Eq.~\eqref{eq:q1} and Eq.~\eqref{eq:q2}, we obtain
\begin{align}
-3\mathcal{U}^{(3)}&=4\Psi_{1}-\frac{3}{8}\eth(h\bar{h})-\frac{3}{2}\bar{h}\bar{Q}^{(1)}\nonumber\\
&=4\Psi_{1}-\frac{3}{8}\eth(h\bar{h})-\frac{3}{2}\bar{h}\eth h
\end{align}
Therefore,
\begin{align}
N=2\Psi_{1}.
\end{align}
Finally, since contracting Eq.~\eqref{eq:bondimomentumaspect} produces
\begin{align}
\widehat{N}=N-u\eth m-\frac{1}{8}\eth(h\bar{h})-\frac{1}{4}\bar{h}\eth h,
\end{align}
we can write the angular momentum aspect as
\begin{align}
\label{eq:angmomaspectfinal}
\widehat{N}&=2\Psi_{1}-u\eth m-\frac{1}{8}\eth(h\bar{h})-\frac{1}{4}\bar{h}\eth h.
\end{align}
As is shown in Secs.~\ref{sec:bmemorycomputation} and~\ref{sec:cmmemorycomputation}, we primarily care about real and imaginary components of this function, which are easily found from Eq.~\eqref{eq:angmomaspectfinal} to be
\begin{subequations}
\begin{align}
\label{eq:reangmom}
\text{Re}(\bar{\eth}\widehat{N})&=\text{Re}\big[2\bar{\eth}\Psi_{1}-\frac{1}{4}\bar{\eth}(\bar{h}\eth h)\big]\nonumber\\
&\phantom{=.}-D^{2}(um+\frac{1}{8}h\bar{h}),\\
\label{eq:imangmom}
\text{Im}(\bar{\eth}\widehat{N})&=\text{Im}\big[2\bar{\eth}\Psi_{1}-\frac{1}{4}\bar{\eth}(\bar{h}\eth h)\big].
\end{align}
\end{subequations}

\section{CM MEMORY}
\label{a:cmmem}
When calculating our expressions for the total memory, we briefly mentioned in Sec.~\ref{sec:cmmemorycomputation} how the electric memory can be seen to contain terms relating to the CM memory.
\begin{figure}
	\label{fig:CMMemoryPlot}
	\centering
	\includegraphics[width=\columnwidth]{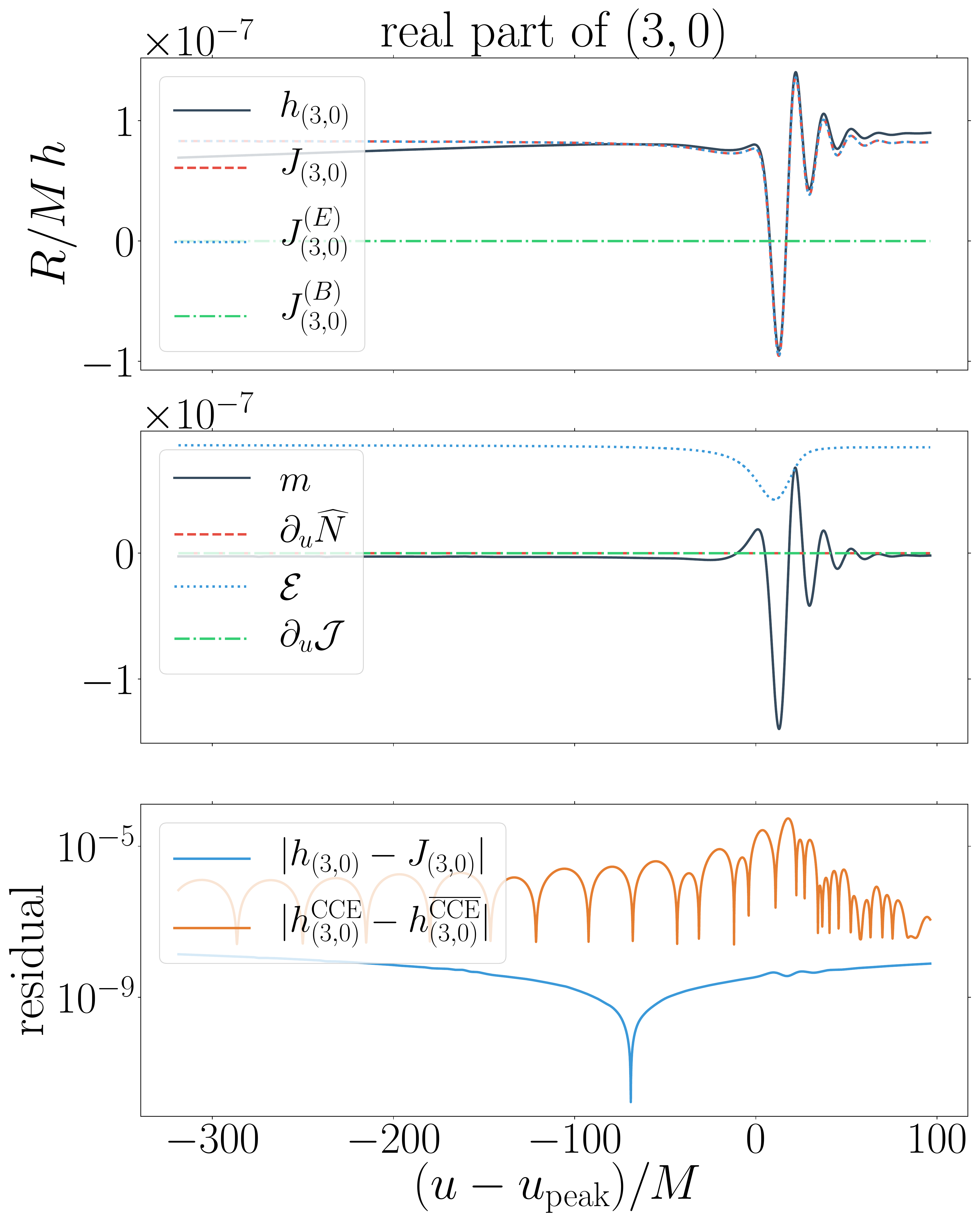}
	\caption{Comparison of the real part of the (3,0) mode of the strain extracted from simulation SXS:BBH:0305 to the strain computed from Eqs.~\eqref{eq:ememoryfinal} and~\eqref{eq:bmemoryfinal}. The top plot shows the extracted strain (black/solid), the strain computed from the BMS flux-balance laws (red/dashed), and its corresponding electric (blue/dotted) and magnetic (green/dashed/dotted) components from Eqs.~\eqref{eq:ememoryfinal} and~\eqref{eq:bmemoryfinal}. The middle plot shows the contributions that come from the mass aspect (black/solid), the angular momentum aspect (red/dashed), the energy flux (blue/dotted), and the angular momentum flux (green/dashed/dotted). We provide an estimate of the strain's corresponding numerical error in the bottom plot.}
\end{figure}
Currently, we are unaware of an explicit formula for the CM memory. For now, though, we present evidence for the CM memory effect in the waveforms produced by numerical relativity. As can be seen in Fig.~\ref{fig:CMMemoryPlot}, while there is no displacement memory present in the mode shown, the energy flux term indicates that when integrated with respect to retarded time this contribution will produce a memory effect, which is exactly the CM memory effect.

\clearpage
\newpage


\bibliography{bibliography}

\end{document}